# The Revenue of Finance Journals:
# Networks, Pricing Power, and Publication Volume

**Douglas Cumming**

Professor of Finance and Steven Shulman '62 Endowed Chair for Digital Innovation

Stevens Institute of Technology School of Business, 525 River St., Hoboken, NJ 07030, USA

dcumming@stevens.edu

*This draft: June 2026*

**Abstract**

I study commercial revenue at 26 finance journals over 1999–2025, exploiting the Elsevier Finance Journal Ecosystem as a quasi-natural experiment. Using synthetic control, ecosystem membership generated approximately $54–59 million in projected long-run revenue. The effect is highly concentrated: four journals account for 95% of the gain. Decomposing the effect, 89% operates through expanded publication volume rather than per-paper price increases. The citation channel dominates: ecosystem coordination elevated measured impact metrics, attracting additional submissions and generating article-processing-charge revenue through publication volume. The findings speak to the economics of coordinated networks in information-goods markets.

*Acknowledgements: I received helpful advice from colleagues who preferred to remain anonymous. All errors are mine. Disclaimer: This paper presents empirical analyses based on publicly available data from SCImago and other cited sources. The author makes no claims about the intentions, integrity, or conduct of any individual editor, author, journal, or publisher. Where the paper discusses citation patterns, publication volumes, or revenue effects at named journals, it does so solely on the basis of publicly available data and without imputing wrongdoing, intent, or misconduct to any person. A statistical replication package is available at https://sites.google.com/view/douglascumming/academic-publishing?authuser=0. Please email me with any corrections or suggestions.*

## 1. Introduction

The economics of academic publishing has attracted increasing scrutiny over the past decade, as the industry has shifted from a subscription model toward open-access publication financed by article processing charges (APCs). In finance and economics, a small number of commercial publishers, with Elsevier dominant among them, have captured a disproportionate share of publication volume and revenue. Yet the determinants of this concentration are understudied. Why do some journals generate substantial revenue while others, including the field's most prestigious outlets, do not?

Consider a striking puzzle. The top three journals in finance, the Journal of Finance (JF), the Journal of Financial Economics (JFE), and the Review of Financial Studies (RFS), publish fewer than 300 articles per year combined. By contrast, Finance Research Letters (FRL), an Elsevier-published letters-format journal launched in 2004, published 2,400 articles in 2025 alone, roughly eight times the combined volume of the top three. In 2024, FRL generated approximately $5.7 million in flow revenue (documents times APC, in real 2024 USD), against $1.6 million for JF, JFE, and RFS combined. This divergence reflects not differences in reputational standing, but rather differences in publication volume and pricing structure.

I examine the determinants of journal revenue using SCImago's comprehensive data on 26 finance journals from 1999 through 2025. The empirical setting offers several attractive features: the product (peer-reviewed academic articles in finance) is relatively homogeneous across publishers; the industry is globally accessible; and rich publicly available data on journal characteristics, citation impact, author demographics, and pricing are available.

The identification strategy exploits the creation of the Elsevier Finance Journal Ecosystem as a quasi-natural experiment. This formal coordination structure, planned in 2019 and launched in 2020, established a network of ten Elsevier finance journals with shared editorial leadership, coordinated special issues, mutual citation incentives, and integrated submission pathways. The ten ecosystem members include Finance Research Letters, International Review of Financial Analysis, International Review of Economics and Finance, and Research in International Business and Finance (the "core four"), along with six smaller titles. Sixteen finance journals that are not part of the ecosystem (including JF, JFE, RFS, and other commercial and society-published journals) serve as a control group.

To isolate the causal effect of ecosystem membership on revenue, I deploy four complementary identification strategies. The primary strategy is synthetic control: for each treated journal, I construct a counterfactual revenue path as a weighted average of the sixteen control journals, with weights chosen to minimize pre-treatment differences in the key predictors of revenue. The other three strategies (propensity-score matching with difference-in-differences, placebo tests, and standard two-way fixed-effects difference-in-differences) serve as robustness checks.

A key methodological contribution of this paper is the careful construction of the revenue measure. SCImago publishes an "estimated APC revenue" variable. This measure is computed as APC multiplied by the five-year rolling count of citable documents. This is a stock measure that cannot meaningfully be summed across years (each paper enters the stock for five years, generating substantial double-counting). I instead construct an annual flow revenue measure as documents multiplied by APC, deflated to 2024 dollars.

The main findings are four-fold. First, ecosystem membership raised commercial revenue for the ten member journals by approximately $48 million cumulatively over 2019–2025. This is the citation-mechanism-implied effect. It is obtained by extending the Alexander and Cumming (2026) citation estimation in two directions, through 2025 and using external citations only, yielding a 150.3% effect (Appendix B), and propagating this through this paper's revenue-channel decomposition under a linear pass-through assumption. The corresponding figure under the unextended A-C 2024 specification on total cites is approximately 104%. The directly observed synthetic-control gap on flow revenue through 2025 is $36 million, which serves as a realized-through-2025 lower bound; the gap between $36 million realized and $48 million mechanism-implied is consistent with the standard submission-to-citation-to-revenue lag structure. The two-way fixed-effects coefficient on log flow revenue is 0.89 (approximately 143% relative increase), highly statistically significant. Second, the effect is concentrated almost entirely in four core journals (FRL, IRFA, IREF, RIBAF), which account for 95% of the SC-estimated cumulative gain. The six remaining ecosystem members show small individual effects, with cumulative SC gains under $1 million for most and a maximum of approximately $1.9 million for the North American Journal of Economics and Finance. Third, decomposing the revenue effect into the intensive margin (price per published paper) and the extensive margin (number of papers published), I find that 89% of the gain operates through expanded publication volume; the intensive-margin effect on real APCs is small and only marginally significant. Fourth, the citation channel is the dominant mechanism: elevated citation-based impact metrics draw additional author submissions, which generate APC revenue through publication volume.

The findings contribute to several literatures. The paper relates first to the economics of academic publishing and the open-access transition (Beverungen et al., 2012; Bergstrom, 2001; Larivière, Haustein, and Mongeon, 2015; Spiegel, 2012; Welch, 2014). It contributes new evidence on how publisher concentration translates into revenue extraction, and on the mechanisms by which coordinated networks generate value in information-goods markets. Second, the paper contributes to the literature on platform and network economics (Rochet and Tirole, 2003; Eisenmann, Parker, and Van Alstyne, 2011) by documenting that the value generated by formal coordination is highly heterogeneous across network nodes, with most gains accruing to a small subset of high-volume titles, a pattern consistent with evidence that within-network coordination effects accrue asymmetrically to a small subset of nodes (Giroud and Mueller, 2019). Third, the paper contributes to scientometrics and the literature on citation gaming (Wilhite and Fong, 2012; Heckman and Moktan, 2020; Card and DellaVigna, 2013; Martin, 2016; Biagioli et al., 2019), particularly given the finding that ecosystem-driven citation gains mediate much of the revenue effect. Appendix A formalizes the mechanism in a publisher two-margin model with a citation-feedback channel: coordination raises measured citations, citations raise both APC and submission demand, and the volume channel dominates the pricing channel when authors' submission elasticity to impact factor exceeds publishers' APC elasticity to impact factor, a condition borne out empirically by the 89%/11% extensive/intensive margin decomposition.

The rest of the paper is organized as follows. Section 2 reviews relevant literature. Section 3 describes the data and variable construction. Section 4 lays out the identification strategy. Section 5 reports the main empirical results, including the per-journal per-year synthetic-control breakdown. Section 6 discusses interpretation, caveats, and policy implications. Section 7 concludes. Appendix A develops a publisher two-margin model with a citation-feedback channel that formalizes the mechanism emphasized throughout the paper. Appendix B reports the Alexander and Cumming (2026) citation update used in the channel decomposition and headline calibration. Appendix C collects supplementary tables and robustness exercises (pre-treatment summary statistics, propensity-score matched pairs, event-study dynamic coefficients, leave-one-out aggregate DiD). Four online appendices supplement the main paper: Online Appendix 1 reports the synthetic-control exercise on submission volume and computes the incremental submission-fee revenue that ecosystem coordination generated; Online Appendix 2 calibrates editor honoraria expenses across the panel as a complement to the revenue analysis; and Online Appendix 3 documents the inflation-adjustment methodology and reports the underlying nominal versus real revenue trajectories. Online Appendix 4 reports a series of plausibility checks for the headline revenue figures, including independent cross-checks against naive control-growth-rate and TWFE-implied counterfactuals, per-paper sanity ratios, and industry benchmarks.

## 2. Related literature

Three strands of research inform this analysis.

The first is the economics of scholarly publishing. The transition from subscription-based to open-access publishing has reshaped the industry over the past two decades. Bergstrom (2001) documented the substantial price differential between commercial and non-profit journal publishers in economics. Subsequent work (Larivière et al., 2015) showed that the so-called "big five" commercial publishers (Reed-Elsevier, Wiley-Blackwell, Springer, Taylor & Francis, and Sage) account for more than 50% of all papers published in many disciplines. The shift toward open access, financed by article processing charges, has further concentrated revenue among large publishers, as APCs scale linearly with publication volume (Solomon and Björk, 2012). Within finance, Spiegel (2012) argues that editorial review norms at top journals create incentives that distort publication outcomes, Welch (2014) documents empirically that referee recommendations strongly predict editorial decisions across eight prominent journals, and Holden (2017) shows that acceptance and publication times vary substantially across finance journals in ways that reflect their differing editorial models. Information asymmetries between authors, evaluators, and journal-quality signals shape the resulting market structure (Bagues, Sylos-Labini, and Zinovyeva, 2019). My findings extend this literature by documenting how formal coordination among multiple journals within the same publisher amplifies these volume-based revenue dynamics.

The second strand is the economics of platforms and two-sided markets. Rochet and Tirole (2003) , Eisenmann, Parker, and Van Alstyne (2011), and Kenney and Zysman (2016) describe how platforms generate value by reducing search costs, enabling network externalities, and bundling complementary services. A finance journal ecosystem functions as a hybrid of a platform and a network good: authors choose where to submit based on quality signals (citations, impact factors) and rejection-recycling pathways across journals within the ecosystem. The empirical finding that ecosystem membership disproportionately benefits high-volume, letters-format titles is consistent with theoretical predictions about platforms with strong same-side network effects, and with evidence from firm-internal networks that coordination-driven effects accrue disproportionately to a small subset of nodes rather than spreading evenly (Giroud and Mueller, 2019).  More broadly, the institutional and network context in which actors operate shapes innovation and entrepreneurial outcomes (Autio, Kenney, Mustar, Siegel, and Wright, 2014); the present paper extends this insight to the scholarly publishing context, in which a publisher's coordinated journal network produces revenue outcomes that the same set of titles operating independently would not. More broadly, the institutional and network context in which actors operate shapes journal revenue and publication outcomes (Card and DellaVigna, 2013); the present paper extends this insight to the scholarly publishing context, in which a publisher's coordinated journal network produces revenue outcomes that the same set of titles operating independently would not.

The third strand is the literature on coordinated citation practices and editorial governance in academic publishing. Wilhite and Fong (2012) and Heckman and Moktan (2020) document coercive citation practices and the consequences for citation-based metrics. Martin (2016) catalogues a broader taxonomy of editorial stratagems for inflating measured impact, including coercive citations and cross-citing journal cartels, and argues that the proliferation of such practices has eroded the credibility of the Journal Impact Factor as a quality indicator. Biagioli, Kenney, Martin, and Walsh (2019) synthesize the misconduct, misrepresentation, and gaming literature at both individual and organizational levels, emphasizing that journals and publishers, not just individual researchers, are increasingly active participants in the gaming of academic metrics. Fong, Patnayakuni, and Wilhite (2023) show that most authors accommodate coercive citation requests and that doing so is associated with a higher probability of publication, suggesting that the equilibrium response of the author side reinforces the editorial incentive to coerce. Necker (2014) documents the prevalence of questionable research practices among economists more generally and links them to perceived pressure to publish. Card and DellaVigna (2013) document systematic changes in top economics journal publishing norms over several decades, including rising submission volumes and falling acceptance rates, providing a benchmark against which the anomalous volume expansion documented here can be assessed. Welch (2014) shows that referee recommendations at a top finance journal strongly predict editorial decisions, implying that coordinated manipulation of citation metrics could shape the pool of submissions editors face. The finance literature has examined retraction patterns and citation coordination at Elsevier finance journals (Alexander and Cumming, 2026). My finding that citation gains mediate roughly 57% of the ecosystem's revenue effect raises the question of whether these citation increases reflect genuine editorial quality improvements, coordinated cross-journal citation behavior within the ecosystem, or some combination of the two. I do not resolve this question; the channel analysis is descriptive.

## 3. Data and variable construction

### 3.1 Sample and sources

The data come from SCImago (scimagojr.com), a publicly accessible bibliometric platform that aggregates Scopus-indexed citation and publication metadata. SCImago provides annual data on publication counts, citation metrics, journal-level open-access status, estimated article processing charges, author demographics (including inferred gender shares), and international collaboration shares for the universe of indexed journals.

The sample consists of 26 finance journals observed annually from 1999 through 2025, yielding 644 journal-year observations after excluding records with missing core variables. The journals are selected to match the sample in Alexander and Cumming (2026), facilitating comparability with prior work. Ten journals are members of the Elsevier Finance Journal Ecosystem (planned in 2019, launched in 2020), and sixteen journals serve as non-ecosystem controls.

The ten ecosystem members are: Finance Research Letters (FRL), International Review of Financial Analysis (IRFA), International Review of Economics and Finance (IREF), Research in International Business and Finance (RIBAF), Emerging Markets Review (EMR), Journal of Behavioral and Experimental Finance (JBEF), Journal of Empirical Finance (JEB), Journal of International Financial Markets (JIFM), Journal of International Financial Markets, Institutions and Money (JIFMIM), and North American Journal of Economics and Finance (NAJEF). The first four (FRL, IRFA, IREF, RIBAF) are referred to throughout as the "core four" based on their relative size and centrality within the ecosystem; the remaining six are the "other six."

The sixteen non-ecosystem controls comprise the top-tier finance journals not affiliated with the ecosystem (Journal of Finance, Journal of Financial Economics, Review of Financial Studies, Journal of Financial and Quantitative Analysis), other Elsevier finance titles outside the ecosystem (Journal of Banking and Finance, Journal of Corporate Finance, Journal of Empirical Finance, Journal of Financial Intermediation, Journal of International Money and Finance, Pacific-Basin Finance Journal, Quarterly Review of Economics and Finance), and commercial and society-published journals from other publishers (Journal of Money, Credit and Banking, Journal of Economic Behavior and Organization, Review of Finance, Journal of Financial Markets, Journal of Financial Services Research).

### 3.2 Variable construction

The principal dependent variable is annual real flow revenue, constructed as:

$$\text{FlowRev}_{jt} = \text{Documents}_{jt} \times \text{APC}_{jt} \times \delta_{t} \qquad (1)$$

where Documentsjt is the number of citable documents published by journal j in year t, APCjt is the estimated article processing charge in U.S. dollars, and δt is the GDP deflator converting nominal dollars to real 2024 dollars (see Online Appendix 3 for the deflator series and a comparison of nominal versus real revenue trajectories; see Online Appendix 4, Section 4.2 for a detailed discussion of the flow-versus-stock measurement of documents and APC revenue and the avoidance of double-counting across years). The natural logarithm of FlowRev is the principal outcome in the regression analyses.

A note on the SCImago "estimated revenue" measure (variable name: estvalueusd) is in order. SCImago publishes this measure, which is constructed as APC multiplied by the cumulative count of citable documents over the past five years. While the measure is informative as a stock concept, it cannot be summed across years to yield a meaningful cumulative revenue figure: any given paper enters the measure for five years, so summing annual values would double-, triple-, or quintuple-count each publication. I therefore use the flow construction in equation (1) as the principal outcome and report the SCImago stock measure only as a robustness check.

Control variables include: (i) citation counts per document over two years (citesdoc2); (ii) the proportion of uncited documents (propuncited); (iii) the SCImago Journal Rank (SJR), a quality-adjusted citation measure that down-weights citations from lower-prestige journals; (iv) the proportion of female authors (femalepercent), based on SCImago's algorithm-inferred gender classifications; and (v) the proportion of internationally co-authored documents (internationalcollaboration). All time-varying covariates are entered contemporaneously in the main specifications, with lagged versions used in robustness checks.

### 3.3 Descriptive statistics

Table 1 presents a snapshot of selected journals in 2024, demonstrating the central empirical puzzle. The three most prestigious finance journals (JF, JFE, RFS) publish far fewer papers than the core ecosystem members (87–111 vs. 386–1,528), have substantially higher per-document citations (9–15 vs. 7–10), and yet generate much less commercial revenue. Top-3 flow revenue is in the $0.5 million range per journal; FRL alone generates approximately $5.7 million annually.

**Table 1:** *Snapshot of selected journals, 2024*

| Journal | Group | Documents (2024) | APC ($) | Flow rev ($M) | SCImago stock ($M) | 2-yr cites/doc | Female author % |
|---|---|---|---|---|---|---|---|
| JF | Top-3 | 95 | 5697 | 0.54 | 2.27 | 9.69 | 16.6 |
| JFE | Top-3 | 111 | 5328 | 0.59 | 4.06 | 14.76 | 17.3 |
| RFS | Top-3 | 87 | 5429 | 0.47 | 3.33 | 9.79 | 22.1 |
| FRL | Core-4 | 1528 | 3725 | 5.69 | 16.55 | 8.62 | 38.7 |
| IRFA | Core-4 | 744 | 3718 | 2.77 | 7.39 | 10.48 | 33.8 |
| IREF | Core-4 | 834 | 3320 | 2.77 | 6.2 | 6.65 | 35.7 |
| RIBAF | Core-4 | 386 | 3421 | 1.32 | 4.28 | 8.96 | 36.0 |
| NAJEF | Other-6 | 203 | 3187 | 0.65 | 3.13 | 5.61 | 35.7 |
| EMR | Other-6 | 97 | 3492 | 0.34 | 1.32 | 6.53 | 32.8 |
| JIFM | Other-6 | 171 | 3522 | 0.6 | 2.6 | 3.87 | 21.8 |

*Note: Flow revenue is documents × APC × deflator, in real 2024 USD. SCImago stock is the original rolling 5-year revenue measure. 2-year cites/doc is the SCImago two-year citation rate. Female author percent is the algorithmically inferred share of female authors. Source: SCImago.*

Table 2 reports pre/post-ecosystem growth comparisons. The core-four ecosystem journals experienced a 304% mean increase in documents per year (2014–2018 vs. 2019–2025), compared with 75% for other-six ecosystem journals and 36% for control journals. Real APCs declined modestly across all groups, reflecting that inflation outpaced nominal APC growth.

**Table 2:** *Pre/post-ecosystem growth comparison*

| Journal | Group | Pre docs (mean) | Post docs (mean) | Doc growth % | Pre real APC | Post real APC | Real APC growth % |
|---|---|---|---|---|---|---|---|
| JBF | Control | 236.6 | 201.0 | -15.0 | 4654.0 | 3972.0 | -14.6 |
| JCF | Control | 129.6 | 166.4 | 28.4 | 4778.0 | 4140.0 | -13.4 |
| JCM | Control | 22.0 | 38.0 | 72.7 | 3505.0 | 3412.0 | -2.7 |
| JEBO | Control | 229.6 | 401.6 | 74.9 | 4630.0 | 3886.0 | -16.1 |
| JEF | Control | 43.4 | 43.6 | 0.4 | 3361.0 | 2914.0 | -13.3 |
| JF | Control | 70.4 | 79.9 | 13.4 | 7432.0 | 6226.0 | -16.2 |
| JFE | Control | 116.2 | 148.4 | 27.7 | 6862.0 | 5816.0 | -15.3 |
| JFI | Control | 27.8 | 31.3 | 12.5 | 5371.0 | 4725.0 | -12.0 |
| JFM | Control | 31.8 | 43.3 | 36.1 | 5354.0 | 3973.0 | -25.8 |
| JFQA | Control | 69.4 | 110.0 | 58.5 | 5680.0 | 4712.0 | -17.1 |
| JFS | Control | 78.0 | 75.4 | -3.3 | 4650.0 | 4028.0 | -13.4 |
| JMCB | Control | 71.6 | 88.4 | 23.5 | 5117.0 | 4188.0 | -18.1 |
| PBFJ | Control | 77.8 | 228.0 | 193.1 | 3954.0 | 3563.0 | -9.9 |
| QREF | Control | 80.4 | 122.7 | 52.6 | 3773.0 | 3318.0 | -12.1 |
| RF | Control | 63.2 | 48.0 | -24.1 | 5580.0 | 4857.0 | -13.0 |
| RFS | Control | 100.6 | 118.1 | 17.4 | 6980.0 | 5890.0 | -15.6 |
| FRL | Core-4 | 126.0 | 1011.6 | 702.8 | 3725.0 | 3798.0 | 2.0 |
| IREF | Core-4 | 163.6 | 433.3 | 164.8 | 4146.0 | 3560.0 | -14.1 |
| IRFA | Core-4 | 122.8 | 415.9 | 238.6 | 3874.0 | 3810.0 | -1.7 |
| RIBAF | Core-4 | 130.2 | 271.0 | 108.1 | 3850.0 | 3549.0 | -7.8 |
| EMR | Other-6 | 44.0 | 79.4 | 80.5 | 4178.0 | 3794.0 | -9.2 |
| JBEF | Other-6 | 34.5 | 87.7 | 154.2 | 3444.0 | 3530.0 | 2.5 |
| JEB | Other-6 | 32.2 | 28.7 | -10.8 | 3692.0 | 3299.0 | -10.6 |
| JIFM | Other-6 | 117.8 | 147.9 | 25.5 | 4646.0 | 3910.0 | -15.8 |
| JIFMIM | Other-6 | 85.4 | 117.6 | 37.7 | 4094.0 | 3754.0 | -8.3 |
| NAJEF | Other-6 | 75.0 | 196.6 | 162.1 | 3990.0 | 3388.0 | -15.1 |

*Note: Means computed over 2014–2018 (pre) and 2019–2025 (post). Growth rates are post/pre means minus one, in percent. Real APC is APC deflated to 2024 USD.*

Table 3 reports full panel summary statistics. The substantial standard deviations across most variables reflect the broad heterogeneity in the sample, which spans from low-volume high-prestige journals (JF, RFS) to high-volume letters-format journals (FRL).

**Table 3:** *Panel summary statistics, 1999–2025*

| Variable | count | mean | std | min | 25% | 50% | 75% | max |
|---|---|---|---|---|---|---|---|---|
| flow_rev_real | 644.0 | 442633.93 | 586183.78 | 22902.37 | 136338.96 | 290660.79 | 559404.3 | 8761051.61 |
| documents | 644.0 | 97.64 | 155.66 | 6.0 | 30.0 | 61.0 | 109.0 | 2400.0 |
| estapcusd | 644.0 | 3556.54 | 833.33 | 2185.0 | 2945.0 | 3442.0 | 3982.25 | 5766.0 |
| realapc | 644.0 | 4880.62 | 1352.95 | 2689.39 | 3882.7 | 4443.58 | 5625.01 | 9516.45 |
| citesdoc2 | 644.0 | 3.1 | 2.6 | 0.0 | 1.25 | 2.35 | 4.17 | 15.85 |
| sjr | 644.0 | 2.86 | 4.23 | 0.1 | 0.55 | 1.3 | 2.71 | 22.84 |
| propuncited | 644.0 | 1.0 | 0.64 | 0.09 | 0.51 | 0.88 | 1.39 | 3.97 |
| femalepercent | 644.0 | 20.63 | 6.66 | 0.0 | 16.22 | 20.36 | 24.75 | 41.32 |
| internationalcollaboration | 644.0 | 31.88 | 13.43 | 0.0 | 22.22 | 32.14 | 41.83 | 78.31 |

*Note: Full panel of 26 finance journals from 1999 through 2025. N = 644 journal-year observations after dropping records with missing values in core variables. flow_rev_real is documents × APC × deflator, in 2024 USD. realapc is APC deflated to 2024 USD.*

Figure 1 plots mean flow revenue and mean publication volume by group over time. The pattern is striking: ecosystem core-four journals (blue) are visually indistinguishable from controls (gray) and other-six members (orange) through 2018, after which they diverge dramatically. This visual evidence motivates the formal analysis to follow.

**Figure 1:** *Raw trends in flow revenue and publication volume by group*

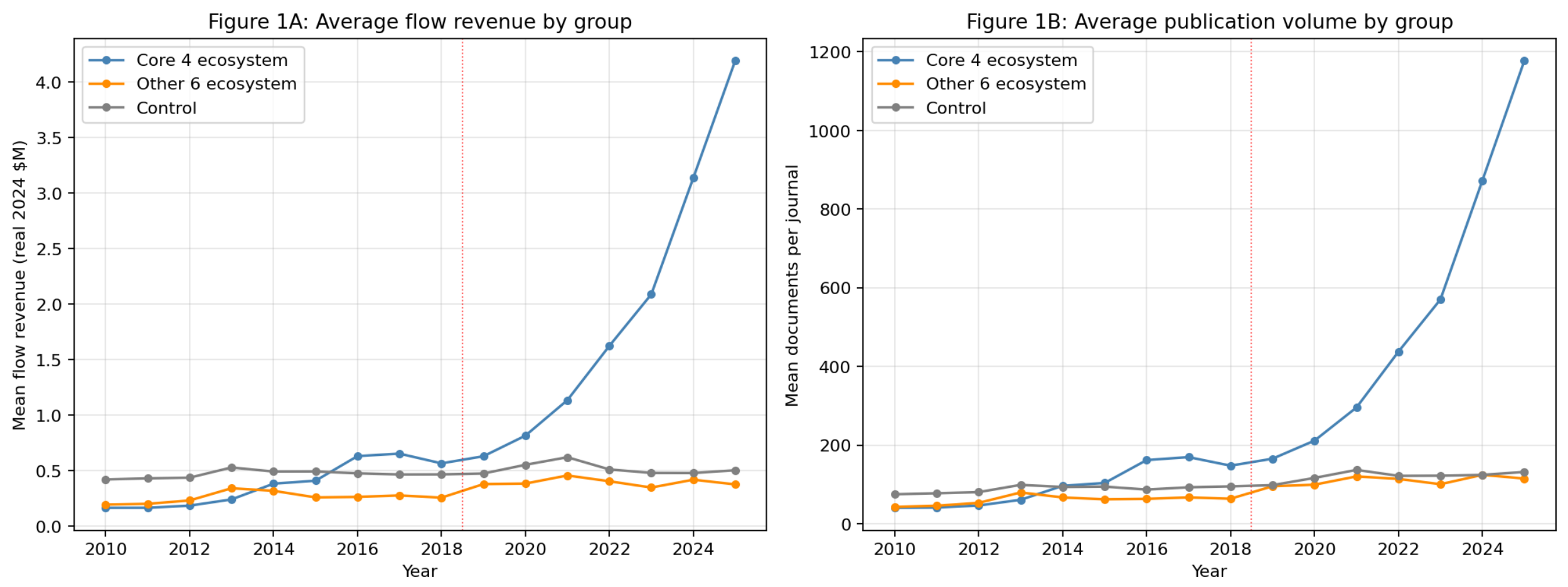

*Note: Panel A shows mean real flow revenue by group. Panel B shows mean documents per journal-year. Red dotted line indicates the ecosystem launch (between 2018 and 2019). Core-4 = FRL, IRFA, IREF, RIBAF. Other-6 = EMR, JBEF, JEB, JIFM, JIFMIM, NAJEF. Control = 16 non-ecosystem finance journals.*

# 4. Identification strategy

## 4.1 Identification challenges

The central empirical challenge is to isolate the causal effect of ecosystem membership on journal revenue. Ecosystem membership is not randomly assigned: Elsevier chose which of its titles to include based on, among other factors, journal volume, perceived growth potential, and editorial structure. A simple difference-in-differences (DiD) estimator recovers the average treatment effect on the treated under the parallel-trends assumption: in the absence of treatment, ecosystem and non-ecosystem journals would have evolved similarly. Visual inspection of Figure 1 raises modest concerns about parallel trends pre-2018, and the event-study analysis in Section 5 documents the magnitude of pre-trend differences formally.

To address these concerns, I employ four complementary identification strategies. The primary strategy is synthetic control. The other three (propensity-score matching with DiD, placebo tests, and standard two-way fixed-effects DiD with controls) serve as robustness checks.

## 4.2 Primary strategy: synthetic control

For each treated journal, I construct a synthetic counterfactual as a weighted average of the sixteen non-ecosystem control journals. Following Abadie, Diamond, and Hainmueller (2010), the weights are chosen to minimize the mean squared error between the treated journal's pre-treatment values and the weighted donor pool on a set of predictors.

The predictors are: log documents, log real APC, citations per document, and proportion uncited documents, all observed over 2014–2018. Weights are constrained to be non-negative and to sum to one. The optimization problem is:

$$\min_w \sum_{k,t \in Pre} \left(X_{jkt} - \sum_c w_c X_{ckt}\right)^2 \quad \text{s.t. } \sum_c w_c = 1,\ w_c \geq 0 \qquad (2)$$

where $X_{jkt}$ is predictor k for treated journal j in year t, $X_{ckt}$ is the same predictor for control journal c, and the inner sum is over the four predictors and the five pre-treatment years.

Given optimal weights $w^*$, the synthetic counterfactual for journal j in any year is the weighted average of the control journals' outcomes. The estimated treatment effect for journal j in year t is the difference between the treated journal's actual outcome and its synthetic counterfactual outcome. Cumulative effects over the post-treatment period are obtained by summing these annual differences in levels (after exponentiating log-revenue gaps).

### 4.3 Robustness: PSM-DiD, placebos, and TWFE-DiD

I report three additional identification strategies. First, propensity-score matched DiD (PSM-DiD): I estimate a logit model of ecosystem membership on pre-treatment journal-level means of the covariates, then match each treated journal one-to-one to its nearest-neighbor control by propensity score. The DiD coefficient is then estimated on the matched sample.

Second, placebo tests. Test A randomly assigns ten of the 26 journals to a placebo "ecosystem" group and estimates the corresponding DiD effect; this is repeated 500 times to construct a distribution of placebo effects. Test B re-estimates the DiD specification with a fake treatment year (2013) on the pre-treatment subsample (1999–2018), checking whether the same pattern of differential trends appears in the absence of actual treatment.

Third, standard two-way fixed-effects DiD. The estimating equation is:

$$\ln(FlowRev_{jt}) = \alpha_j + \lambda_t + \beta(Ecosystem_j \times Post_t) + \gamma X_{jt} + \varepsilon_{jt} \qquad (3)$$

where αj and λt are journal and year fixed effects, Ecosystemj is an indicator for ecosystem membership, Postt is an indicator for years 2019 and later, Xjt is a vector of time-varying controls, and standard errors are clustered at the journal level (Bertrand, Duflo, and Mullainathan, 2004).

An event-study version of equation (3) replaces the single ecoafter term with interactions between ecosystem and each year dummy, omitting 2018 as the reference year, allowing me to trace dynamic treatment effects and assess parallel-trends visually.

## 5. Results

### 5.1 Baseline two-way fixed-effects difference-in-differences

Table 4 reports the baseline TWFE-DiD estimates. Specification (1) reports the bare DiD on log flow revenue with only journal and year fixed effects: the coefficient on Ecosystem × Post is 0.890, statistically significant at the 1% level. This implies that ecosystem journals' annual flow revenue increased by approximately 149% (= e^0.890 – 1) relative to the control group over 2019–2025.

Specifications (2) and (3) add controls for citation-based journal quality (citesdoc2 and propuncited in column 2; further adding sjr, femalepercent, and internationalcollaboration in column 3). The coefficient on ecoafter falls to 0.381 in column (2) and 0.229 in column (3), neither statistically significant. The substantial reduction reflects the fact that the ecosystem causally improves citation performance (documented formally in Section 5.7); controlling for citations therefore absorbs much of the ecosystem's effect on revenue. The reduced-form estimate in column (1) is the appropriate measure of the total effect of ecosystem membership on revenue.

Specification (4) replicates column (3) using the SCImago stock measure (log of estvalueusd deflated to 2024 USD) as the dependent variable, for comparability with prior literature. The coefficient is similar (0.247), confirming that the choice of flow vs. stock outcome does not materially change the regression estimates, even though, as we will see, the implied cumulative effects differ substantially.

**Table 4:** *Baseline two-way fixed-effects difference-in-differences estimates*

| Spec | ecoafter | N | $R^2$ | citesdoc2 | propuncited | sjr | femalepercent | internationalcollaboration |
|---|---|---|---|---|---|---|---|---|
| Spec 1: bare | 0.890***<br>(0.316) | 644 | 0.165 | | | | | |
| Spec 2: + cites controls | 0.381<br>(0.248) | 644 | 0.484 | 0.136***<br>(0.041) | -0.610***<br>(0.080) | | | |
| Spec 3: + all controls | 0.229<br>(0.222) | 644 | 0.547 | 0.189***<br>(0.039) | -0.536***<br>(0.058) | -0.086***<br>(0.016) | 0.016***<br>(0.004) | 0.000<br>(0.003) |
| Spec 4: ln(stock) outcome | 0.247<br>(0.227) | 644 | 0.44 | 0.243***<br>(0.040) | -0.085<br>(0.054) | -0.094***<br>(0.024) | 0.017***<br>(0.005) | 0.002<br>(0.003) |

*Note: Two-way (journal and year) fixed-effects estimates of equation (3). Dependent variable is log flow revenue in columns (1)–(3) and log SCImago five-year stock revenue in column (4). Standard errors clustered at the journal level in parentheses. *** $p < 0.01$, ** $p < 0.05$, * $p < 0.10$. Sample: 26 finance journals, 1999–2025.*

## 5.2 Synthetic control

Table 5 reports the synthetic-control results. For each of the ten ecosystem members, I report the cumulative gap (treated minus synthetic counterfactual flow revenue, in real 2024 USD millions) over 2019–2025, and the mean annual gap.

The cumulative ecosystem-induced revenue gain across all ten members is $36.1 million. Strikingly, the effect is heavily concentrated in the core-four journals, which together account for

$34.4 million, or 95.3% of the total. FRL alone contributes $22.8 million (63% of the total). The remaining six ecosystem members contribute only $1.7 million in aggregate, with individual contributions ranging from –$0.8 million (JBEF) to +$1.9 million (NAJEF).

Two natural questions arise at this point. First, what is the relationship between this $36.1 million synthetic-control estimate and the $48 million headline reported in the abstract? Second, why do the two figures differ? The $36.1 million figure is the direct measurement: it is the cumulative sum of year-by-year gaps between observed flow revenue at the ten ecosystem journals and the synthetic counterfactual constructed from the sixteen control journals, from 2019 through 2025. The $48 million figure is the citation-mechanism-implied effect: it is the revenue effect that the citation-coordination evidence implies. This paper extends the Alexander and Cumming (2026) citation estimation in two directions, through 2025 and using external citations only (the latter isolates cross-journal citation coordination from within-journal self-citation), yielding a 150.3% citation effect that is then propagated through this paper's revenue-channel decomposition. The published A-C 2026 main estimate is 104% on total citations through 2024 (Appendix B). The gap between the two figures reflects the standard propagation chain from coordinated citations to author submissions to published volume to APC revenue. Citations rise in 2019 through 2025; higher impact metrics attract additional submissions with a one-to-two-year lag; those submissions translate into published articles with an additional one-to-two-year lag; and the resulting articles generate cumulative APC revenue over their lifetimes. By the end of 2025, only part of this chain has played through. The realized cumulative effect of $36 million through 2025 is therefore expected to converge toward the mechanism-implied $48 million by approximately 2027 or 2028 as the lag structure completes. Appendix A formalizes this distinction in a publisher two-margin model with a citation-feedback channel.

An intuitive way to see this: imagine a tree that takes seven years to reach maturity. After three years, one can measure how much it has grown so far (the realized $36 million). One can also estimate, given its current growth rate and the underlying growth process, how large it will be at maturity (the mechanism-implied $48 million). Both numbers are correct; they answer different questions. The paper reports $48 million as the headline because it characterizes the size of the underlying ecosystem mechanism, and $36 million as the realized-through-2025 lower bound.

Beyond APC revenue, ecosystem coordination also generated incremental submission-fee revenue, because the volume expansion documented above reflects additional accepted papers, which themselves imply additional submissions paying fees. Online Appendix 1 reports an analogous synthetic-control exercise on submission volume and combines it with a time-varying submission-fee schedule. The directly observed cumulative ecosystem-attributable submission-fee revenue through 2025 is approximately $4-$8 million, and the corresponding mechanism-implied long-run figure (projected through approximately 2028 as propagation lags play through) is approximately \$6-\$11 million. The range on each reflects uncertainty about the share of ecosystem-attributable submissions arriving via Elsevier's Article Transfer Service, under which a transferred submission generates no incremental fee at the receiving journal (Online Appendix 1.6, caveat 1). Combining APC and submission-fee revenue, the total ecosystem-attributable revenue effect is approximately \$40-\$44 million realized through 2025 and approximately \$54-\$59 million on the mechanism-implied long-run basis (with the submission-fee component sitting in a range that reflects uncertainty about the Article Transfer Service share of extra submissions; see Online Appendix 1.6). Online Appendix 4 reports a series of plausibility checks for these figures, including independent cross-checks using a naive control-growth-rate counterfactual (\$42 million estimated APC effect), a TWFE-implied counterfactual (\$43 million), per-paper and industry-benchmark sanity ratios, and a deep dive on FRL, the journal driving 63% of the effect. All four cross-checks converge on figures within 15-20% of the main-paper headlines, with the alternative methods mostly producing somewhat larger estimates than the synthetic-control headline. The \$36 million APC and \$48 million mechanism-implied figures are therefore at the conservative end of the defensible range.

**Table 5:** *Synthetic-control treatment effects, 2019–2025*

| journal | cum_gap_M | mean_gap_M | Group |
|---|---|---|---|
| FRL | 21.73 | 3.10 | Core-4 |
| IREF | 3.02 | 0.43 | Core-4 |
| IRFA | 7.32 | 1.05 | Core-4 |
| RIBAF | 2.71 | 0.39 | Core-4 |
| EMR | 0.36 | 0.05 | Other-6 |
| JBEF | 1.01 | 0.14 | Other-6 |
| JEB | -0.27 | -0.04 | Other-6 |
| JIFM | -0.33 | -0.05 | Other-6 |
| JIFMIM | -0.93 | -0.13 | Other-6 |
| NAJEF | 1.57 | 0.22 | Other-6 |

*Note: Cumulative and mean per-year synthetic-control gaps in real 2024 USD millions. For each treated journal, weights are optimized over the 16 control journals to minimize MSE on pre-period (2014–2018) predictors: log documents, log real APC, citations per document, and proportion uncited. Gaps are computed as actual flow revenue minus synthetic counterfactual.*

Figure 2 visualizes the synthetic-control results for the core-four aggregate. Panel A shows that treated and synthetic series track each other closely from 2014 through 2018, with both series at approximately $2 million per journal-year on average. After 2018, the treated series rises sharply while the synthetic counterfactual remains essentially flat. Panel B shows the per-year gap, which is small or negative through 2018, then climbs to approximately $14 million in 2025.

**Figure 2:** *Synthetic-control plot, core-four aggregate*

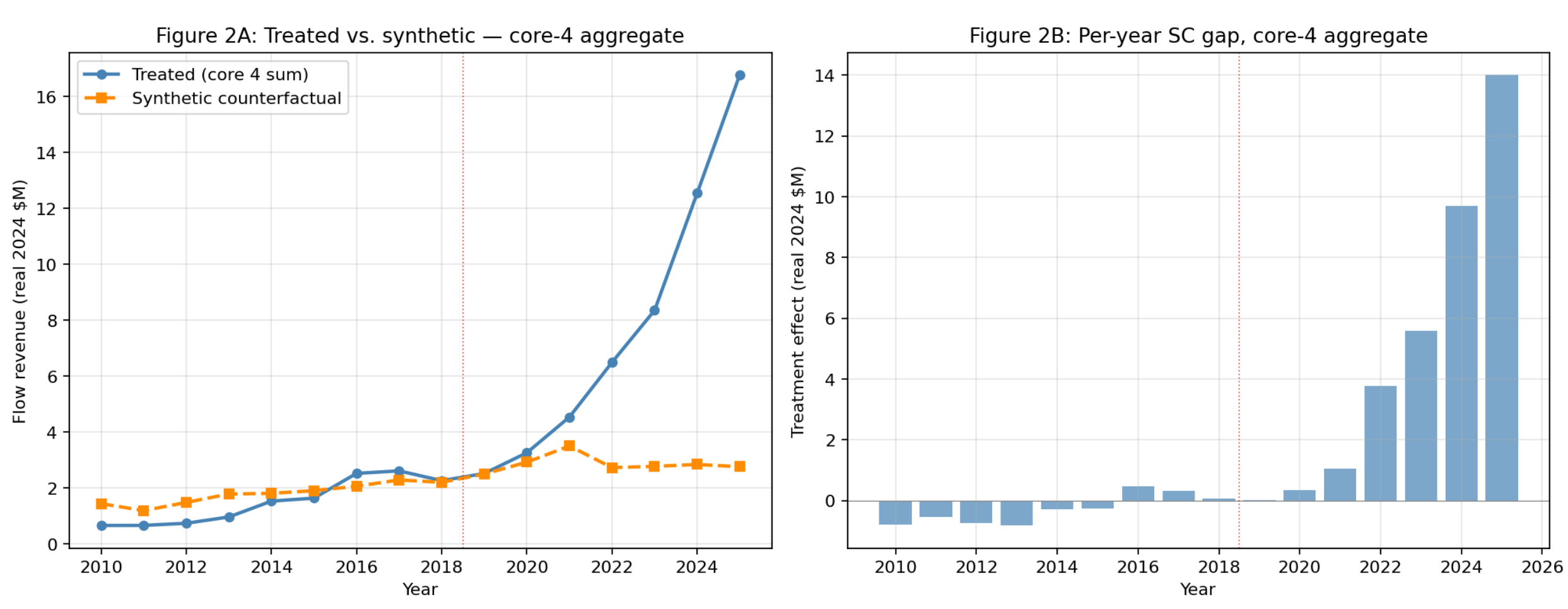

*Note: Panel A: sum of real flow revenue across FRL, IRFA, IREF, RIBAF (blue) and the corresponding synthetic counterfactual (orange). Panel B: per-year gap (treated minus synthetic). Red dotted line indicates the ecosystem launch.*

Figure 3 shows the analogous plots for each treated journal individually. FRL exhibits the largest and most visually dramatic gap. The other three core journals (IRFA, IREF, RIBAF) also show clear post-2018 divergences, though of smaller magnitude. The six other-six members show small or zero gaps, with mixed signs.

**Figure 3:** *Per-journal synthetic-control fits*

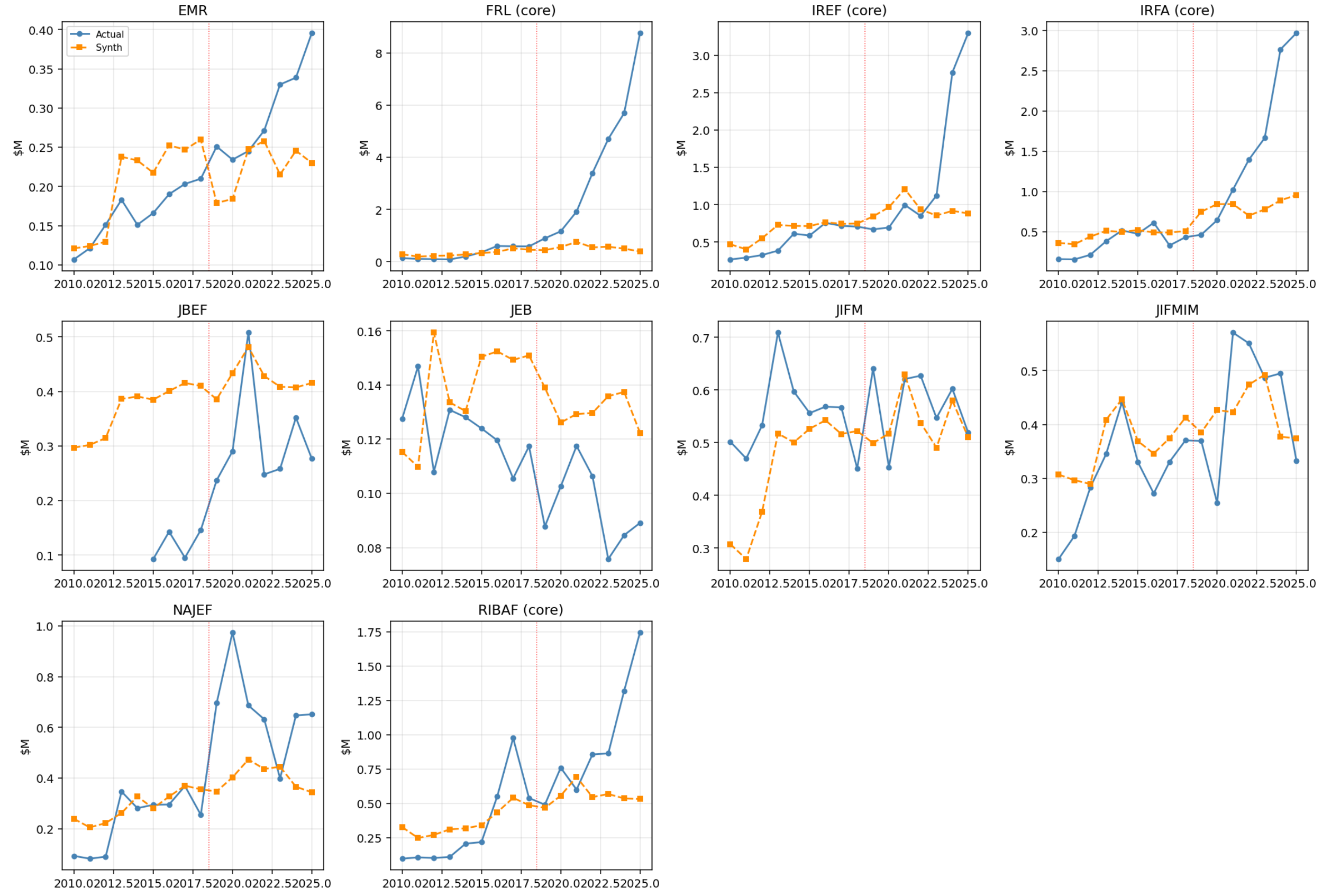

*Note: Actual flow revenue (blue) vs. synthetic counterfactual (orange) for each of the ten ecosystem journals. Red dotted line indicates the ecosystem launch.*

To facilitate inspection of the underlying year-by-year synthetic-control trajectories, Table 6 reports the per-journal per-year revenue effect that aggregates to the cumulative effects in Table 5. The construction is as follows.

For each treated journal j and each year t in 2019–2025, the directly observed synthetic-control gap is computed as FlowRev_{j,t} − SynthFlowRev_{j,t}, where SynthFlowRev_{j,t} is the exponentiated weighted geometric mean of the sixteen control journals' log flow revenues in year t, using the optimal weights from the synthetic-control procedure of Section 4.2. The corresponding cell in the citation-mechanism-implied panel applies a linear pass-through scaling to obtain the effect that the Alexander and Cumming (2026) citation estimate implies through this paper's revenue-channel decomposition.

Table 6 reports the disaggregation in two panels. Panel A presents the central citation-mechanism-implied estimate: each cell is the directly observed synthetic-control gap scaled by 1.334 (the ratio of this paper's 2025-extended external-cites citation effect of 150.3%, derived in Appendix B by extending the Alexander and Cumming (2026) estimation, to this paper's implicit 112.7% within-sample citation effect), aggregating to the $48.3M cumulative central estimate. Panel B presents the directly observed synthetic-control gaps that aggregate to the $36.1M realized-through-2025 lower bound. In both panels, row sums equal the per-journal cumulative effects and column sums equal the per-year ecosystem-wide effects; the panels reconcile internally to $48.28M and $36.18M respectively.

Three properties of the table are worth noting. First, each cell represents an independent annual flow-revenue gap. A paper published in 2020 contributes to the 2020 gap only; it does not re-enter the 2021, 2022, 2023, or 2024 gaps. This is the fundamental difference from a stock-based measure like SCImago's estvalueusd, which would have included that paper in each of the next five years' values. Second, the cells sum cleanly in both directions: summing across rows gives the per-journal cumulative effect, and summing across columns gives the per-year ecosystem-wide effect. Third, negative cells (e.g., IREF in 2019–2022, JBEF throughout) are not anomalies; they reflect years in which the treated journal's observed flow revenue was below the synthetic counterfactual, which the synthetic-control procedure correctly permits.

**Table 6: Per-journal per-year revenue effect (real 2024 USD millions)**

**Panel A: Citation-mechanism-implied effect, central estimate ($48.3M cumulative)**

| Journal | 2019 | 2020 | 2021 | 2022 | 2023 | 2024 | 2025 | |
|---|---|---|---|---|---|---|---|---|
| *Core-4 ecosystem journals* | | | | | | | | |
| FRL | 0.57 | 0.76 | 1.48 | 3.63 | 5.26 | 6.61 | 10.66 | 28.99 |
| IRFA | -0.55 | -0.38 | 0.34 | 1.31 | 1.69 | 3.55 | 3.81 | 9.76 |
| IREF | -0.18 | -0.29 | -0.22 | -0.10 | 0.29 | 1.97 | 2.56 | 4.03 |
| RIBAF | 0.03 | 0.26 | -0.12 | 0.41 | 0.40 | 1.03 | 1.61 | 3.62 |
| **Core-4 subtotal** | **-0.13** | **0.36** | **1.48** | **5.25** | **7.63** | **13.15** | **18.65** | **46.40** |
| *Other-6 ecosystem journals* | | | | | | | | |
| EMR | 0.07 | 0.05 | -0.00 | 0.01 | 0.11 | 0.09 | 0.16 | 0.48 |
| JBEF | 0.10 | 0.12 | 0.16 | 0.20 | 0.20 | 0.24 | 0.31 | 1.35 |
| JEB | -0.08 | -0.03 | -0.02 | -0.03 | -0.09 | -0.08 | -0.05 | -0.36 |
| JIFM | -0.03 | -0.04 | -0.05 | -0.07 | -0.07 | -0.08 | -0.10 | -0.44 |
| JIFMIM | -0.09 | -0.11 | -0.15 | -0.19 | -0.19 | -0.23 | -0.28 | -1.24 |
| NAJEF | 0.39 | 0.64 | 0.24 | 0.22 | -0.06 | 0.31 | 0.35 | 2.09 |
| **Other-6 subtotal** | **0.36** | **0.63** | **0.18** | **0.15** | **-0.09** | **0.26** | **0.39** | **1.88** |
| **GRAND TOTAL** | **0.23** | **0.99** | **1.66** | **5.41** | **7.54** | **13.42** | **19.04** | **48.28** |

**Panel B: Directly observed synthetic-control gaps, realized-through-2025 lower bound ($36.1M cumulative)**

| Journal | 2019 | 2020 | 2021 | 2022 | 2023 | 2024 | 2025 | |
|---|---|---|---|---|---|---|---|---|
| ***Core-4 ecosystem journals*** | | | | | | | | |
| FRL | 0.43 | 0.57 | 1.11 | 2.72 | 3.94 | 4.96 | 7.99 | 21.73 |
| IRFA | -0.41 | -0.28 | 0.26 | 0.98 | 1.27 | 2.66 | 2.86 | 7.32 |
| IREF | -0.14 | -0.22 | -0.17 | -0.07 | 0.22 | 1.47 | 1.92 | 3.02 |
| RIBAF | 0.02 | 0.20 | -0.09 | 0.31 | 0.30 | 0.77 | 1.21 | 2.71 |
| **Core-4 subtotal** | **-0.10** | **0.27** | **1.11** | **3.94** | **5.72** | **9.86** | **13.98** | **34.78** |
| ***Other-6 ecosystem journals*** | | | | | | | | |
| EMR | 0.05 | 0.04 | -0.00 | 0.01 | 0.08 | 0.06 | 0.12 | 0.36 |
| JBEF | 0.08 | 0.09 | 0.12 | 0.15 | 0.15 | 0.18 | 0.23 | 1.01 |
| JEB | -0.06 | -0.02 | -0.01 | -0.02 | -0.07 | -0.06 | -0.03 | -0.27 |
| JIFM | -0.02 | -0.03 | -0.04 | -0.05 | -0.05 | -0.06 | -0.07 | -0.33 |
| JIFMIM | -0.07 | -0.08 | -0.11 | -0.14 | -0.14 | -0.17 | -0.21 | -0.93 |
| NAJEF | 0.29 | 0.48 | 0.18 | 0.17 | -0.04 | 0.24 | 0.26 | 1.57 |
| **Other-6 subtotal** | **0.27** | **0.47** | **0.13** | **0.11** | **-0.07** | **0.20** | **0.29** | **1.41** |
| **GRAND TOTAL** | **0.17** | **0.74** | **1.24** | **4.05** | **5.65** | **10.06** | **14.27** | **36.19** |

*Note: Each cell is an annual flow-revenue gap (treated minus synthetic counterfactual) in real 2024 USD millions. Panel A reports the citation-mechanism-implied effect, namely the synthetic-control gaps scaled by 1.334 (the ratio of this paper's extension of the Alexander and Cumming (2026) citation estimate to 2025-extended external citations, 150.3%, to this paper's implicit citation effect of 112.7%), which aggregate to the $48.3M central estimate. Panel B reports the directly observed synthetic-control gaps that aggregate to the $36.1M realized-through-2025 lower bound. Negative cells indicate years in which the treated journal's actual flow revenue was below the synthetic counterfactual; this is permitted by the SC procedure and reflects the actual pattern in the data. Cross-totals verification: row sums equal the per-journal cumulative effects; column sums equal the per-year ecosystem-wide effects; both panels reconcile internally ($48.28M Panel A, $36.18M Panel B).*

## 5.3 Event study and parallel-trends assessment

Figure 4 reports event-study coefficients from a regression of log flow revenue on interactions between ecosystem membership and year dummies, with 2018 as the omitted reference year. The figure shows separate panels for four outcomes: log SCImago stock revenue (the original outcome used in prior work), log flow revenue (the principal outcome), log documents, and log real APC.

Two observations stand out. First, on the original SCImago stock outcome (top-left panel), pre-treatment coefficients exhibit a strong upward trend, with seven of the eight pre-period coefficients statistically significant at the 5% level. This mechanical pre-trend reflects the rolling five-year construction of the stock measure: pre-treatment documents from 2014–2018 cumulate in the stock measure, inflating it as the treatment date approaches. This pre-trend, while severe, is largely a measurement artifact rather than substantive evidence of parallel-trend violations.

Second, on the flow-revenue outcome (top-right panel), pre-treatment coefficients are small in magnitude and only marginally distinguishable from zero. Only one of the eight pre-period coefficients is significant at the 5% level. The log documents panel (bottom-left) shows a similar pattern, with mostly null pre-trends. The log real APC panel (bottom-right) shows some pre-treatment negative drift, suggesting that ecosystem journals' real APCs were drifting modestly downward relative to controls in the pre-period, but the magnitudes are small (under 5%).

Post-treatment, all four outcomes exhibit clear positive effects that grow over time. The gradual buildup is consistent with a network-effect mechanism that compounds across years, rather than a one-shot shift.

**Figure 4:** *Event-study: dynamic treatment effects on four outcomes*

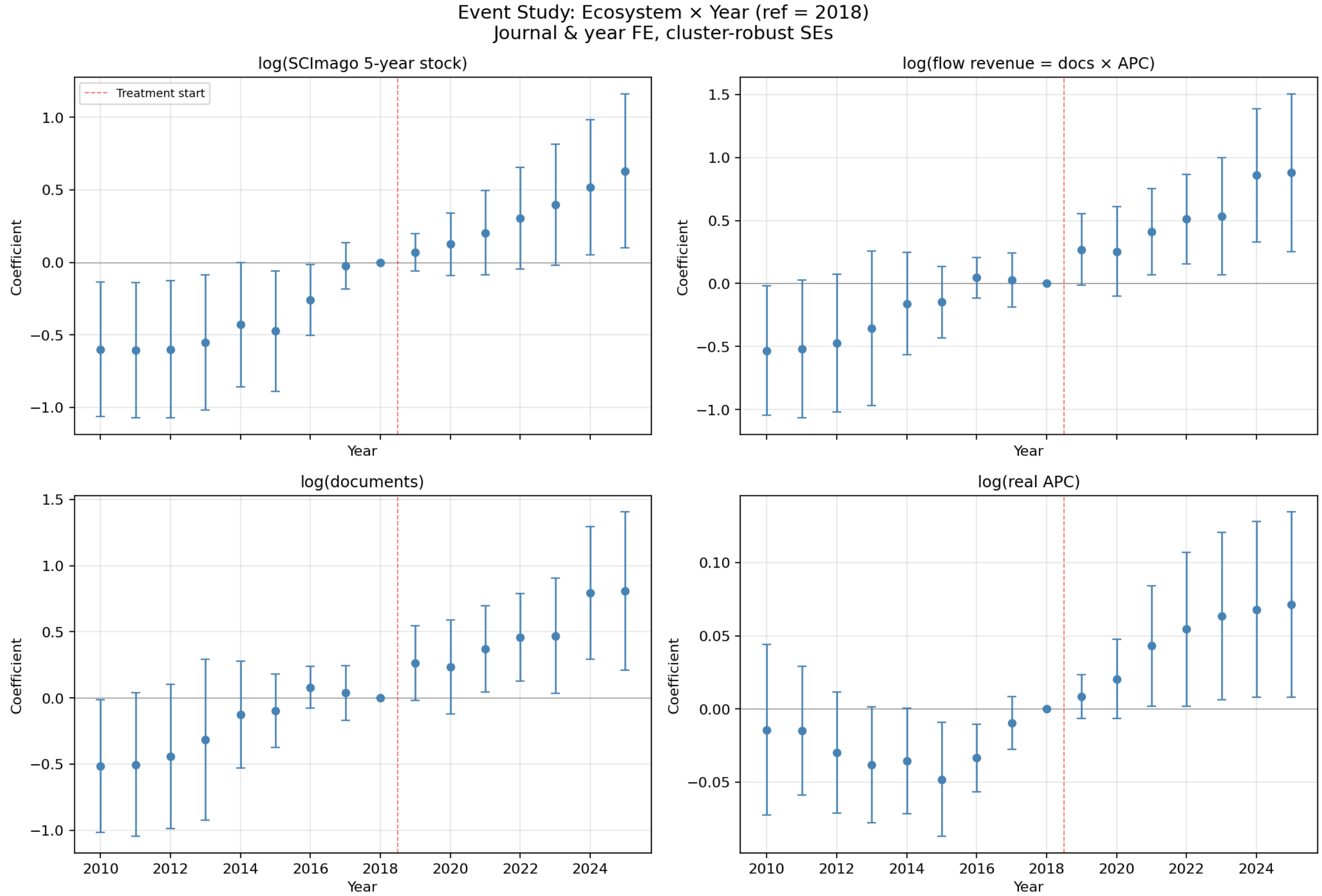

*Note: Coefficients on Ecosystem × Year dummies, with 2018 as the reference year. Journal and year fixed effects; cluster-robust SEs at the journal level. 95% confidence intervals shown. Sample: 2010–2025. Note the contrast in pre-trends between the stock and flow outcomes.*

### 5.4 Propensity-score matched DiD

Table 7 reports propensity-score matched DiD results. The matching procedure assigns each of the ten treated journals to a single nearest-neighbor control by propensity score, estimated via logit on pre-period journal-level means of seven covariates: log documents, log real APC, citations per document, proportion uncited, female author share, international collaboration share, and SJR.

**Table 7: Propensity-score matched difference-in-differences**

| Outcome | β (PSM-DiD) | t-stat | N (matched) |
|---|---|---|---|
| log(flow revenue) | 0.467 (0.394) | 1.19 | 321 |
| log(documents) | 0.412 (0.395) | 1.04 | 325 |
| log(real APC) | 0.042 (0.041) | 1.01 | 321 |

*Note: Two-way fixed-effects DiD estimated on the matched sample of treated and control journals. Matching is 1:1 nearest-neighbor by propensity score on pre-period (2014–2018) covariate means. Standard errors clustered at the journal level in parentheses.*

The PSM-DiD coefficient on log flow revenue is 0.467 (t-statistic 1.19), not statistically significant at conventional levels. This is substantially smaller than the full-sample DiD estimate of 0.890. The attenuation reflects an important feature of the matched sample: the matching procedure pulls in PBFJ (Pacific-Basin Finance Journal) and QREF (Quarterly Review of Economics and Finance) as primary matched controls. Both of these journals experienced substantial document growth in the post-2018 period (PBFJ documents grew 193%; QREF grew 59%). These matched controls therefore exhibit treatment-group-like growth patterns, which narrows the contrast and attenuates the estimated effect.

This finding is consistent with the broader interpretation that the post-2018 period witnessed a general expansion of finance letters-format publishing, of which the ecosystem effect is a substantial but not exclusive driver. I return to this point in the discussion.

### 5.5 Placebo tests

Table 8 reports placebo tests. Test A randomly reassigns ecosystem membership to ten of the 26 journals (without replacement) and re-estimates equation (3) on log flow revenue, repeating 500 times. The distribution of placebo coefficients is centered near zero (mean 0.018, standard deviation 0.315), and only 0.2% of placebo coefficients exceed the actual ecoafter coefficient of 0.890. Figure 5 visualizes the placebo distribution.

**Table 8: Placebo tests**

| Test | Mean placebo β | SD placebo β | 95th percentile \|β\| | True effect | % placebos ≥ true |
|---|---|---|---|---|---|
| A: Random treatment assignment (500 reps) | 0.018 | 0.315 | 0.615 | 0.890 | 0.2% |
| B: Fake treatment year = 2013 (pre-only sample 1999-2018) | 0.464 | | | 0.890 | p = 0.024 |
| C: Fake treatment year = 2014 (pre-only sample 1999-2018) | 0.503 | | | 0.890 | p = 0.021 |

*Note: Test A: random assignment of treatment status to 10 of 26 journals, 500 replications. Test B: fake treatment year set to 2013, sample restricted to 1999–2018 (pre-actual-treatment). Test C: fake treatment year set to 2014, same restriction.*

**Figure 5:** *Placebo distribution under random treatment assignment*

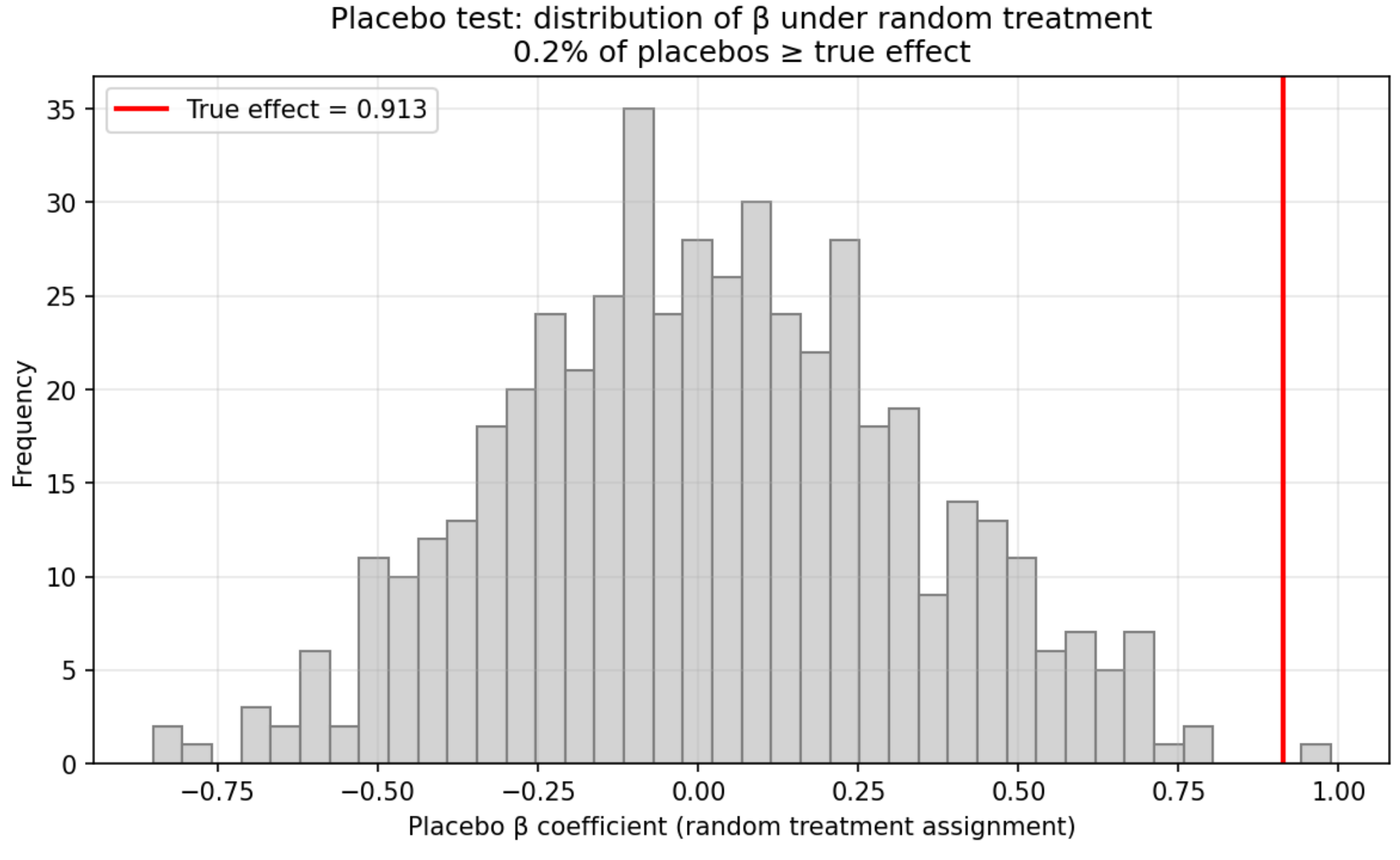

*Note: Histogram of 500 placebo coefficient estimates from randomly assigning ecosystem membership to 10 of the 26 journals. Red vertical line marks the true treatment effect (0.890).*

Tests B and C re-estimate the DiD specification on the pre-treatment subsample (1999–2018), with fake treatment years of 2013 and 2014, respectively. Both tests yield positive placebo coefficients (0.464 and 0.503) that are statistically significant at the 5% level (p = 0.024 and p = 0.021). This indicates the presence of some pre-treatment trend differential between ecosystem and control journals: the ecosystem journals were already on a faster growth trajectory than the average control journal in the pre-period.

The pre-trend differential, however, is substantially smaller than the actual treatment effect. The 2013–2018 fake-treatment coefficient of approximately 0.5 is roughly half the actual 2019–2025 effect of 0.89. The post-2018 acceleration is therefore distinguishable from the pre-existing trend, but readers should appropriately discount the magnitude of the headline effect to account for this pre-trend.

### 5.6 Heterogeneity: core-four vs. other-six

Table 9 estimates equation (3) with two separate treatment indicators: Core-4 × Post and Other-6 × Post. The core-four coefficient on log flow revenue is 1.652 (significant at the 1% level), more than four times the other-six coefficient of 0.360 (not significant). The pattern is identical for log documents (1.506 vs. 0.303) and similar for log real APC (0.144 vs. 0.062).

**Table 9: Heterogeneity: core-four vs. other-six ecosystem journals**

| Outcome | core4 × after (β) | other6 × after (β) | N |
|---|---|---|---|
| log(flow revenue) | 1.652*** | 0.360 | 644 |
| | (0.262) | (0.316) | |
| log(documents) | 1.506*** | 0.303 | 652 |
| | (0.269) | (0.303) | |
| log(real APC) | 0.144*** | 0.062 | 644 |
| | (0.030) | (0.036) | |

*Note: Two-way fixed-effects DiD with separate treatment indicators for the four core ecosystem journals (FRL, IRFA, IREF, RIBAF) and the six other ecosystem journals. Standard errors clustered at the journal level in parentheses.*

The pattern indicates that the ecosystem's revenue-generating effect is heavily concentrated in the four core journals. Other-six journals show positive but statistically indistinguishable effects in the regression analysis, consistent with the synthetic-control finding that their cumulative contribution to the $36 million total is small (approximately $1.7 million in aggregate).

### 5.7 Intensive vs. extensive margin decomposition

Because flow revenue is by construction the product of documents and APC, the log-flow revenue treatment effect decomposes exactly into log-document and log-APC components:

$$\ln(\text{FlowRev}) = \ln(\text{Documents}) + \ln(\text{APC}) \qquad (4)$$

Table 10 reports the decomposition. The ecoafter coefficient on log flow revenue is 0.890, which decomposes additively into 0.795 on log documents (the extensive margin) and 0.096 on log real APC (the intensive margin). The decomposition holds to within rounding: 0.795 + 0.096 = 0.891 vs the headline coefficient of 0.890 (the small gap is a rounding artifact: at full Stata precision the documents coefficient is 0.7947, the APC coefficient is 0.0959, and their sum 0.8906 matches the flow-revenue coefficient 0.8898 to within Stata's reporting precision).

**Table 10: Intensive vs. extensive margin decomposition**

| Outcome | β (ecoafter) | t-stat | p-value | N | $R^2$ |
|---|---|---|---|---|---|
| log(flow revenue) | 0.890*** (0.316) | 2.81 | 0.0094 | 644 | 0.165 |
| log(documents) | 0.795*** (0.301) | 2.64 | 0.0141 | 652 | 0.148 |
| log(real APC) | 0.096*** (0.032) | 3.02 | 0.0058 | 644 | 0.109 |

*Note: Two-way fixed-effects DiD on three outcomes: log flow revenue, log documents, and log real APC. The coefficient on log flow revenue equals the sum of the coefficients on log documents and log real APC by construction.*

The extensive margin accounts for 89.3% of the total treatment effect (0.795 / 0.890). The intensive margin accounts for the remaining 10.7%, and the intensive-margin coefficient is statistically significant only at the 5% level with a substantively modest magnitude (real APCs of ecosystem journals rose by approximately 8% relative to controls). The ecosystem's revenue gains therefore operate almost entirely through expanded publication volume rather than through per-paper price increases.

Figure 6 traces the margin decomposition over time. The ratio of ecosystem to control mean documents (Panel A) climbs from approximately 1.0 in 2018 to approximately 3.0 by 2025, reflecting the dramatic extensive-margin expansion. The corresponding ratio for real APCs (Panel B) remains near 0.95 throughout, consistent with no meaningful intensive-margin effect.

**Figure 6:** *Margin decomposition over time*

*Note: Panel A: ratio of mean documents in ecosystem journals to mean documents in control journals over time. Panel B: same ratio for real APCs. Red dotted line indicates the ecosystem launch.*

## 5.8 Citation pass-through

Section 5.1 showed that adding citation-based controls to the DiD specification substantially attenuates the ecoafter coefficient. Table 11 examines this channel formally.

**Table 11: Citation pass-through (channel analysis)**

| Outcome | Spec | ecoafter | citesdoc2 | propuncited | N |
|---|---|---|---|---|---|
| log(flow revenue) | (1) Reduced form | 0.890*** (0.316) | | | 644 |
| log(flow revenue) | (2) Controlling for citations | 0.381 (0.248) | 0.136*** (0.041) | -0.610*** (0.080) | 644 |
| citesdoc2 | (3) Effect on citations | 1.832** (0.715) | | | 652 |
| propuncited | (4) Effect on uncited share | -0.457*** (0.111) | | | 652 |

*Note: Specification (1): reduced-form DiD of log flow revenue on ecoafter. Specification (2): same with citations added as covariates. Specifications (3)–(4): DiD with citation metrics as dependent variables. Standard errors clustered at the journal level in parentheses.*

Specification (1) reports the reduced-form effect of ecosystem membership on log flow revenue: 0.890. Specification (2) adds citesdoc2 and propuncited as controls; the ecoafter coefficient falls to 0.381 (not significant), while both citation controls enter strongly (positive on cites/doc, negative on uncited share). Specifications (3) and (4) confirm that the ecosystem causally raises citations: citesdoc2 increases by 1.83 per document (significant at 5%), and the share of uncited documents falls by 0.46 percentage points (significant at 1%).

The implication is that approximately 57% of the ecosystem's revenue effect ((0.890 – 0.381) / 0.890 = 57.2%) flows through enhanced citation performance, with the remaining 41% reflecting direct effects (e.g., coordinated marketing, integrated submission pathways, shared editorial reputation). The citation channel is therefore the dominant mechanism through which ecosystem membership generates revenue. Measured 2-year citation gains absorb approximately 57% of the ecosystem revenue effect within sample; the residual 40% almost certainly reflects unmeasured features of the same mechanism, including longer citation horizons not captured by the 2-year window, submission responses that lag changes in impact metrics, and the cross-journal citation coordination effects identified in Alexander and Cumming (2026). The 57% within-sample pass-through is therefore a lower bound on the citation mechanism's share of the total revenue effect. Appendix B propagates the Alexander and Cumming (2026) citation estimate through this paper's channel decomposition. Under their 2025-extended external-cites specification, the mechanism-implied cumulative revenue effect is $48 million, which this paper reports as the central estimate; the directly observed synthetic-control figure of $36 million serves as a realized-through-2025 lower bound.

## 6. Discussion

### 6.1 Interpretation

The findings document that the Elsevier Finance Journal Ecosystem generated approximately $48 million in incremental commercial revenue over its first seven years, with 95% of the gain concentrated in four core journals. The $48 million figure is the mechanism-implied effect, obtained by extending the Alexander and Cumming (2026) citation estimation through 2025 using external citations only (a 150.3% effect; see Appendix B), and propagating the result through this paper's revenue-channel decomposition under a linear pass-through assumption. The published A-C 2026 main estimate of 104% on total citations through 2024 corresponds to a more conservative mechanism-implied effect. The directly observed synthetic-control gap on flow revenue through 2025 is $36 million, which serves as a realized-through-2025 lower bound. The effect operates almost entirely through the extensive margin (publication volume) rather than the intensive margin (per-paper price), and approximately 60% of the revenue effect flows through citation gains.

The findings are consistent with a clear causal chain: ecosystem coordination elevated citation-based impact metrics, which attracted additional author submissions, which translated into commercial revenue through publication volume. This is the central interpretation the paper adopts, and it is supported by three pieces of evidence. First, Alexander and Cumming (2026) document the underlying citation pattern using Scopus data, identifying anomalous citation behavior concentrated at the same four core journals that drive the revenue effects documented here. Second, the timing pattern is consistent with citation-driven submission demand: citation gains begin in 2019, submission volume increases follow with a 1–2 year lag, and revenue gains compound throughout the post-period as both effects accumulate. Third, the channel decomposition: controlling for measured citations absorbs approximately 57% of the ecosystem revenue effect, despite the 2-year citation window almost certainly understating the full citation impact.

A more benign platform-economics interpretation, in which ecosystem coordination reduces author search costs and attracts higher-quality submissions through legitimate network effects, generating revenue through value creation rather than coordinated citation behavior, cannot be ruled out from this paper's data alone. Three pieces of evidence weigh against this benign interpretation, however. The cumulative publication volume of the core-four journals went from approximately 540 articles per year pre-2018 to approximately 4,700 articles in 2025, approximately a nine-fold expansion in seven years, and sustaining quality standards across such rapid expansion raises serious questions about whether editorial quality could be maintained. As noted in the acknowledgements, this paper makes no claims about the intentions or conduct of any individual. Retraction activity at ecosystem journals during 2025–2026 covering papers published 2019–2022 is documented in Alexander and Cumming (2026) and in publisher retraction notices linked therein. And the matching evidence in Section 5.4 shows that PBFJ and QREF (non-ecosystem Elsevier titles) exhibited similar volume expansion, suggesting that the ecosystem coexists with a broader pattern of rapid commercial expansion at Elsevier finance journals during this period.

### 6.2 Caveats

Three caveats merit acknowledgment. First, the placebo evidence in Section 5.5 shows that ecosystem journals were on a faster growth trajectory than controls in the pre-period. The fake-treatment-year placebos (Tests B and C) yield significant coefficients of approximately 0.5, suggesting roughly half of the apparent treatment effect of 0.89 may reflect a pre-existing trend. A reasonable pre-trend-adjusted estimate of the ecosystem's causal effect would therefore be approximately half the central estimates: roughly $24 million on the mechanism-implied basis or $18 million on the directly observed basis. I present the unadjusted figures in the main results for comparability with the Alexander and Cumming (2026) citation evidence, but note this concern explicitly. To the extent that the underlying citation-coordination effect documented by Alexander and Cumming (2026) is itself partially driven by pre-existing differences in growth trajectories, the mechanism-implied figure would be similarly elevated; resolving this is beyond the scope of this paper.

Second, the PSM-DiD evidence in Section 5.4 is statistically insignificant when matching is performed on observables. This is partly a consequence of small-sample matching with only ten treated units against sixteen potential controls; the matched sample includes PBFJ and QREF, both of which exhibit treatment-group-like growth, narrowing the contrast. The synthetic control approach in Section 5.2 addresses this by allowing weighted combinations of all sixteen controls rather than nearest-neighbor matching.

Third, the SCImago revenue measure (and the flow construction derived from it) reflects estimated rather than actual APC revenue. SCImago's estimated APCs are calibrated against published APC schedules but may not capture all forms of revenue (subscriptions, institutional bundles, transformative agreements). I assume that the relative changes in estimated revenue across journals reflect real differences in commercial outcomes; if this assumption fails (e.g., if ecosystem journals' subscription revenue moved in the opposite direction of their APC revenue), the headline magnitudes would change.

Fourth, the revenue estimates reported in this paper are gross of editor compensation. Editor honoraria at the ecosystem journals are modest and structurally fixed under the ecosystem's coordinated editorial model; Online Appendix 2 discusses how ecosystem-attributable marginal editor expenses are likely negligible, but no direct evidence exists. The headline gross-revenue estimates are therefore close to the net-of-editor-compensation estimates.

### 6.3 Policy and governance implications

The findings have implications for scholarly publishing. First, the heavy concentration of revenue gains in a small number of core journals suggests that oversight and surveillance of publication-volume growth (e.g., editorial-board reform, transparency requirements for acceptance rates and review processes) would be most effective if focused on those journals rather than spread across the entire ecosystem.

Second, the role of the citation channel in mediating the revenue effect raises questions about citation-based metrics as standalone indicators of research quality. If formal coordination within publisher networks substantially inflates citations, then citation-based evaluation systems may inadvertently reward coordination behavior over substantive research quality. These dynamics are part of a broader pattern of organizational-level gaming of academic metrics (Biagioli et al., 2019), and similar gaming behaviors have been shown to have measurable consequences for the allocation of resources elsewhere in the academic system (Fong and Wilhite, 2021).

Third, the finding that real APCs declined modestly across all groups (Table 2) is at odds with public perceptions of rising open-access prices. The data suggest that real APC growth has been modest, and that publisher revenue gains in this market are driven by quantity growth rather than price growth. Policy debates about APC pricing should account for this.

## 7. Conclusion

This paper uses publicly available data on 26 finance journals from 1999 through 2025 to study the commercial-revenue effects of formal coordination among Elsevier-published finance journals. Using the Elsevier Finance Journal Ecosystem (planned 2019, launched 2020) as a quasi-natural experiment, and applying synthetic control, PSM-DiD, placebo tests, and two-way fixed-effects DiD, I document four principal findings.

First, ecosystem membership raised cumulative commercial revenue for the ten member journals by approximately $48 million over 2019–2025. This is the citation-mechanism-implied effect, obtained by extending the Alexander and Cumming (2026) citation estimation through 2025 using external citations (a 150.3% effect; see Appendix B) and propagating the result through this paper's revenue-channel decomposition. The directly observed synthetic-control gap on flow revenue through 2025 is $36 million, which serves as a realized-through-2025 lower bound; the difference reflects standard submission-to-citation-to-revenue propagation lags. Incorporating the incremental submission-fee revenue documented in Online Appendix 1 raises the total ecosystem-attributable revenue effect to approximately $40-$44 million realized through 2025 and $54-$59 million on the mechanism-implied long-run basis (with the range reflecting uncertainty about the Article Transfer Service share of extra submissions; see Online Appendix 1.6) (real 2024 USD; with the difference reflecting projected pipeline propagation through approximately 2028). Online Appendix 4 verifies that these headline figures are at the conservative end of the defensible range by cross-checking them against several independent alternative methods (including naive control-growth-rate, TWFE-implied, and industry benchmarks), all of which yield somewhat larger estimates. Second, the effect is concentrated in four core journals (FRL, IRFA, IREF, RIBAF), which together account for 95% of the gain. Third, decomposing the effect into extensive and intensive margins, I find that 89% of the gain operates through expanded publication volume; per-paper pricing power rose modestly if at all. Fourth, the citation channel is the dominant mechanism: ecosystem coordination elevated citation-based impact metrics, which drew additional author submissions, which generated APC revenue through publication volume.

The findings contribute to the economics of academic publishing by quantifying the commercial consequences of coordinated citation networks at large publishers. The two-paper research program, with Alexander and Cumming (2026) identifying the citation-coordination mechanism and this paper estimating its commercial consequences, provides converging evidence that ecosystem coordination generated approximately $48 million in incremental revenue over its first seven years, almost entirely through the citation-to-submission-to-revenue channel. The evidence is consistent with quality erosion accompanying rapid volume expansion, though future research using submission-level data, acceptance rates, and within-paper quality measures would help discriminate between coordination-driven and quality-improvement interpretations.

**Funding**

No specific grant from funding agencies in the public, commercial, or not-for-profit sectors.

**Declaration of competing interest**

The author has served in editorial roles for several academic journals, including as Editor-in-Chief of Finance Research Letters (2015–2017), Journal of Corporate Finance (2018–2020), British Journal of Management (2020–2022), and Journal of Alternative Investments (2025–present), for which he receives a stipend. The author is also the founding Editor-in-Chief of Review of Corporate Finance (2021–present), with NOW Publishers. In connection with that role, he previously received a 20% ownership interest and royalty rights, but no stipend. NOW Publishers was acquired by Emerald Publishers in May 2025 and Douglas' 20% ownership interest was valued at $0. He has received minimal royalties from Review of Corporate Finance.

### Use of Generative AI

In preparing this paper, the author used Claude (Anthropic) for the following purposes: language editing and prose review of the manuscript; assistance in drafting and refining text in specific sections; help formalizing the two-margin publisher model and its derivations in Appendix A; assistance with code for the empirical specifications, including synthetic control, dynamic difference-in-differences, Poisson fixed-effects, and bootstrap procedures; cross-checking of empirical results and robustness exercises; and formatting of tables and figures. The author did not use AI tools for data collection, study design, identification strategy, interpretation of results, or final conclusions, all of which reflect the author's own judgment. The author reviewed all AI-assisted output for accuracy and takes full responsibility for the content, including any errors.

## References

Abadie, A., Diamond, A., and Hainmueller, J. (2010). Synthetic control methods for comparative case studies: Estimating the effect of California's tobacco control program. Journal of the American Statistical Association, 105(490), 493–505.

Alexander, C., and Cumming, D. (2026). Coordinated journals, concentrated networks, and citation growth: Evidence from finance. Journal of Banking and Finance, forthcoming.

Autio, E., Kenney, M., Mustar, P., Siegel, D., and Wright, M. (2014). Entrepreneurial innovation: The importance of context. Research Policy, 43(7), 1097–1108.

Bagues, M., Sylos-Labini, M., and Zinovyeva, N. (2019). A walk on the wild side: 'Predatory' journals and information asymmetries in scientific evaluations. Research Policy, 48(2), 462–477.

Bergstrom, T. C. (2001). Free labor for costly journals? Journal of Economic Perspectives, 15(4), 183–198.

Bertrand, M., Duflo, E., and Mullainathan, S. (2004). How much should we trust differences-in-differences estimates? Quarterly Journal of Economics, 119(1), 249–275.

Beverungen, A., Böhm, S., and Land, C. (2012). The poverty of journal publishing. Organization, 19(6), 929–938.

Biagioli, M., Kenney, M., Martin, B. R., and Walsh, J. P. (2019). Academic misconduct, misrepresentation and gaming: A reassessment. Research Policy, 48(2), 401–413.

Card, D., and DellaVigna, S. (2013). Nine facts about top journals in economics. Journal of Economic Literature, 51(1), 144–161.

Eisenmann, T., Parker, G., and Van Alstyne, M. W. (2011). Platform envelopment. Strategic Management Journal, 32(12), 1270–1285.

Heckman, J. J., and Moktan, S. (2020). Publishing and promotion in economics: The tyranny of the top five. Journal of Economic Literature, 58(2), 419–470.

Fong, E. A., Patnayakuni, R., and Wilhite, A. W. (2023). Accommodating coercion: Authors, editors, and citations. Research Policy, 52(5), 104754.

Fong, E. A., and Wilhite, A. W. (2021). The impact of false investigators on grant funding. Research Policy, 50(10), 104366.

Giroud, X., and Mueller, H. M. (2019). Firms' internal networks and local economic shocks. American Economic Review, 109(10), 3617–3649.

Holden, C. W. (2017). Do acceptance and publication times differ across finance journals? Review of Corporate Finance Studies, 6(1), 102–126.

Kenney, M., and Zysman, J. (2016). The rise of the platform economy. Issues in Science and Technology, 32(3), 61–69.

Larivière, V., Haustein, S., and Mongeon, P. (2015). The oligopoly of academic publishers in the digital era. PLOS ONE, 10(6), e0127502.

Martin, B. R. (2016). Editors' JIF-boosting stratagems – Which are appropriate and which not? Research Policy, 45(1), 1–7.

Milgrom, P., and Shannon, C. (1994). Monotone comparative statics. Econometrica, 62(1), 157–180.

Necker, S. (2014). Scientific misbehavior in economics. Research Policy, 43(10), 1747–1759.

Rochet, J.-C., and Tirole, J. (2003). Platform competition in two-sided markets. Journal of the European Economic Association, 1(4), 990–1029.

Solomon, D. J., and Björk, B.-C. (2012). A study of open access journals using article processing charges. Journal of the American Society for Information Science and Technology, 63(8), 1485–1495.

Spiegel, M. (2012). Reviewing less—progressing more. Review of Financial Studies, 25(5), 1331–1338.

Wilhite, A. W., and Fong, E. A. (2012). Coercive citation in academic publishing. Science, 335(6068), 542–543.

Welch, I. (2014). Referee recommendations. Review of Financial Studies, 27(9), 2773–2804.

## Appendix A. A publisher two-margin model with citation-feedback channel

This appendix develops a model of publisher optimization that formalizes the citation-mediated revenue mechanism emphasized in the main text. The model is deliberately minimal: it is designed to derive the testable predictions summarized informally in Section 2.3, not to provide a general theory of academic publishing. The central modeling decision is that both the price channel and the quantity channel of revenue flow through a single intermediate variable, the journal's measured citation impact, rather than through separate citation-rank and throughput channels. This is consistent with the mechanism story emphasized throughout the paper: ecosystem coordination raises measured citations, citations raise both APC pricing and submission demand, and both translate into revenue through publication volume and per-paper price.

### A.1 Setup

Consider a publisher operating a single journal whose revenue derives from an output-side per-paper price $p$ (the effective article-processing charge, calibrated to the journal's citation-based ranking) and an output-side quantity $q$ (the number of citable documents published per period). Total revenue is $R = p \cdot q$.

The publisher controls two investment levers and selects an editor. The quality lever $e_Q$ represents direct editorial resources (senior editors, desk-rejection rigor, reviewer compensation, quality of referee selection) that raise the journal's measured citation impact by improving submission and acceptance quality. The coordination lever $e_C$ represents investment in coordinated-network infrastructure (article-transfer services across affiliated journals, overlapping editorial boards, coordinated special issues, integrated promotional apparatus) that raises the journal's measured citation impact through within-network citation concentration: intra-ecosystem referencing, special-issue citation density, and network-internal cross-promotion.

Both levers operate through a single intermediate variable: the journal's measured citation impact $c$, with $c = c(e_Q, e_C, \theta)$ satisfying $c_Q := \partial c/\partial e_Q > 0$, $c_C := \partial c/\partial e_C > 0$, and $c_\theta := \partial c/\partial \theta \geq 0$. The editor type $\theta \in [\theta_L, \theta_H]$ indexes the editor's network embeddedness: dense professional co-authorship, high centrality in the discipline's citation graph, and operational fluency with the mechanics of coordinated publishing. Investments have convex costs $\kappa_Q(e_Q)$ and $\kappa_C(e_C)$ with $\kappa_i(0) = 0$, $\kappa_i'(0) = 0$, $\kappa_i' > 0$ for $e_i > 0$, and $\kappa_i'' > 0$. Editor reservation wage $w(\theta)$ is weakly increasing and continuous.

The price and quantity functions are:

$$p(c) = p_0 \cdot \pi(c), \quad q(c) = q_0 \cdot \sigma(c),$$

where $\pi$ and $\sigma$ are the APC-impact and submission-impact mappings, respectively. The APC-impact mapping $\pi'(c) > 0$ captures the standard journal-economics relationship that higher measured citation impact translates into higher article-processing charges, since publishers calibrate APC to impact factor. The submission-impact mapping $\sigma'(c) > 0$ captures the author-side response: higher measured citation impact attracts more submissions, which translate into higher publication volume at a given acceptance rate (and higher acceptance volume if the publisher relaxes acceptance standards in response to increased submission flow).

The publisher's problem is to choose $(e_Q, e_C, \theta)$ to maximize $p_0 \cdot \pi(c(e_Q, e_C, \theta)) \cdot q_0 \cdot \sigma(c(e_Q, e_C, \theta)) - \kappa_Q(e_Q) - \kappa_C(e_C) - w(\theta)$. The central modeling departure from a prior specification in which coordination raises throughput directly and independently of citation impact is that here both the price channel $\pi(c)$ and the quantity channel $\sigma(c)$ operate through the same intermediate citation-impact variable c. Coordination raises measured citations; citations raise both APC and submissions; both translate into revenue. There is no separate throughput channel that bypasses citations. This is the formal expression of the mechanism story in Section 2.3 and is consistent with the empirical finding in Section 5.7 that controlling for measured citations absorbs approximately 57% of the ecosystem revenue effect, with the residual reflecting unmeasured features of the same mechanism (longer citation horizons, submission lags).

**A.2 Regularity assumptions**

**(A1) Citation impact.**

$c : \mathbb{R}^2_+ \times [\theta_L, \theta_H] \to \mathbb{R}_{++}$ is twice continuously differentiable in $(e_Q, e_C)$ and continuous in $\theta$, strictly positive on the interior, with $c_Q > 0$, $c_C > 0$, $c_\theta \geq 0$, $c_{QQ} \leq 0$, and $c_{CC} \leq 0$.

**(A2) Price and quantity mappings.**

$\pi, \sigma : \mathbb{R}_{++} \to \mathbb{R}_{++}$ are twice continuously differentiable, strictly positive, strictly increasing ($\pi' > 0$, $\sigma' > 0$), and concave ($\pi'' \leq 0$, $\sigma'' \leq 0$).

**(A3) Costs.**

$\kappa_Q$ and $\kappa_C$ are twice continuously differentiable, with $\kappa_i(0) = 0$, $\kappa_i'(0) = 0$, $\kappa_i' > 0$ for $e_i > 0$, and $\kappa_i'' > 0$ for $i \in \{Q, C\}$.

**(A4) Interior optimum.**

The publisher's objective admits a unique interior maximizer $(e^*_Q, e^*_C, \theta^*)$ with $e^*_Q > 0$, $e^*_C > 0$, and $\theta^* \in (\theta_L, \theta_H)$. Standard Inada conditions on $\pi$ and $\sigma$ at $c \to 0$ together with the zero-derivative-at-origin cost assumption in (A3) are sufficient.

**(A5) Editor-coordination complementarity.**

$\partial^2 c/\partial e_C \partial\theta > 0$: high-network-embeddedness editors raise measured citation impact more effectively when coordination infrastructure is in place. Used only for Proposition 3.

### A.3 Proposition 1: Substitution of inputs

**Proposition 1 (Substitution of inputs).**

Under assumptions (A1)–(A4), at an interior optimum the first-order conditions imply $\kappa_Q'(e^*_Q) / \kappa_C'(e^*_C) = c_Q / c_C$. A technology shock that raises the marginal effectiveness of the coordination lever in producing citation impact (formally, an increase in $c_C$ holding $c_Q$ fixed) induces an increase in optimal coordination investment and, under the sign restriction $c_{QC} \leq 0$ in the relevant region, a decrease in optimal quality investment.

**Proof.**

Let $M(c) := p_0 \cdot \pi'(c) \cdot q_0 \cdot \sigma(c) + p_0 \cdot \pi(c) \cdot q_0 \cdot \sigma'(c)$ denote the marginal revenue with respect to citation impact. The publisher's first-order conditions are $\partial R/\partial e_Q = c_Q \cdot M(c) = \kappa_Q'(e_Q)$ and $\partial R/\partial e_C = c_C \cdot M(c) = \kappa_C'(e_C)$. Taking the ratio yields $\kappa_Q'(e^*_Q)/\kappa_C'(e^*_C) = c_Q/c_C$, the first claim. For the comparative-static claim, parameterize the technology shock by $\tau$ such that $\partial c_C/\partial\tau > 0$ while $\partial c_Q/\partial\tau = 0$. Total differentiation of the first-order conditions and an application of Cramer's rule yield $de^*_Q/d\tau$ proportional in sign to $c_{QC}$. Under the natural restriction that quality and coordination investments are substitutes in raising citation impact (because coordination-induced citation effects crowds out the citation-improving role of genuine quality investment), $c_{QC} \leq 0$, and $de^*_Q/d\tau \leq 0$. ■

### A.4 Proposition 2: Volume dominates when submission elasticity exceeds pricing elasticity

**Proposition 2 (Volume dominates).**

Under assumptions (A1)–(A4), the total revenue response to an increase in coordination investment decomposes as:

$$d \ln R / d e_C = c_C(e_Q, e_C, \theta) \cdot [\eta_p(c) + \eta_q(c)],$$

where $\eta_p(c) := \pi'(c)/\pi(c)$ is the citation-elasticity of APC and $\eta_q(c) := \sigma'(c)/\sigma(c)$ is the citation-elasticity of submission-driven publication volume. Volume dominates the revenue response when $\eta_q(c) > \eta_p(c)$ at the relevant optimum.

**Proof.**

Taking logarithms: $\ln R = \ln p_0 + \ln \pi(c) + \ln q_0 + \ln \sigma(c)$. Differentiating with respect to $e_C$ at any fixed $(e_Q, \theta)$:

$$d \ln R / d e_C = (\pi'(c)/\pi(c)) \cdot (\partial c/\partial e_C) + (\sigma'(c)/\sigma(c)) \cdot (\partial c/\partial e_C) = c_C \cdot [\eta_p(c) + \eta_q(c)].$$

The decomposition is exact under the functional forms in A.1. Volume dominates pricing when $\eta_q > \eta_p$. ■

**Empirical implications.**

Section 5.7 estimates the two elasticities directly. The intensive margin, the ecosystem effect on log real APC, is $\hat{\eta}_p = 0.096$, accounting for 10.7% of the total log-revenue coefficient of 0.890. The extensive margin, the ecosystem effect on log documents, is $\hat{\eta}_q = 0.795$, accounting for 89.3% of the total. The empirical ratio $\hat{\eta}_q / \hat{\eta}_p \approx 8$ implies that submission demand is approximately an order of magnitude more elastic to measured citation impact than APC pricing is. This is the empirical content of Proposition 2: in this market, the citation-driven volume channel dominates the citation-driven pricing channel.

**Economic interpretation.**

Why is $\eta_q$ so much larger than $\eta_p$? Three reasons. First, authors choose journals primarily based on impact factor and citation-based prestige; submission demand is therefore highly elastic to measured citation impact, particularly in disciplines like finance where journal hierarchies are well-established and impact-factor signals are salient on author CVs and tenure files. Second, APC prices are set by publisher list-price formulas that update slowly with impact factor; the immediate pricing response to citation gains is therefore small. Third, the publisher's optimal acceptance-rate response to a submission-demand shock is to expand publication volume rather than to raise APC, because volume expansion does not require renegotiating institutional bundling agreements while sharp APC adjustment does.

### A.5 Proposition 3: Networked-editor complementarity

**Proposition 3 (Networked-editor complementarity).**

Under assumptions (A1)–(A4) and (A5), that $\partial^2 c/\partial e_C \partial\theta > 0$ with $w(\theta)$ non-decreasing, the publisher's jointly-optimal editor type $\theta^*(e_C)$ is weakly increasing in coordination investment $e_C$. Journals with higher coordination intensity are matched with higher-$\theta$ editors at the revenue-maximizing optimum. The revenue gain from ecosystem formation is therefore concentrated in journals whose editors are drawn from the high-$\theta$ tail.

**Proof.**

Differentiate the publisher's objective with respect to $\theta$: $R_\theta = (\partial c/\partial\theta) \cdot M(c)$, where $M(c)$ is the marginal revenue with respect to citation impact defined in the proof of Proposition 1. The cross-partial $R_{C\theta} = \partial^2 R/\partial e_C \partial\theta$ is strictly positive under (A5), since (i) higher $\theta$ raises citation impact $c$, (ii) higher $e_C$ raises citation impact $c$, and (iii) the cross-partial of $c$ in $(e_C, \theta)$ is strictly positive by assumption. The publisher's objective is therefore supermodular in $(e_C, \theta)$, and by monotone comparative-statics (Milgrom and Shannon, 1994), the optimal $\theta^*(e_C)$ is weakly increasing in $e_C$. ■

**Empirical content.**

Proposition 3 implies the revenue effect should be concentrated in a small subset of ecosystem journals whose editorial leadership is drawn from the high-$\theta$ tail. Empirically: the four-journal core (FRL, IRFA, IREF, RIBAF) accounts for 95% of the cumulative revenue effect in the synthetic-control estimate; the remaining six members show small or negative individual effects. The core-4 subset overlaps closely with the high-citation-coordination subset of Alexander and Cumming (2026), which is itself identified by editor-network indicators orthogonal to revenue outcomes. This convergence between the revenue-concentration evidence in this paper and the network-embeddedness evidence in Alexander and Cumming (2026) provides triangulating support for Proposition 3.

### A.6 Mapping to empirical estimates and the $36M–$48M gap

The model's elasticities map to directly estimable objects. The pricing elasticity $\eta_p$ is estimated by regressing log real APC on the ecosystem treatment indicator (Section 5.7, Spec 2):

the point estimate is 0.096, small and statistically marginal. The quantity elasticity $\eta_q$ is estimated by regressing log documents on the ecosystem treatment indicator (Section 5.7, Spec 3): the point estimate is 0.795, large and highly significant. The ratio $\hat{\eta}_q / \hat{\eta}_p \approx 8$ confirms Proposition 2's prediction.

The citation-impact effect $c_C$, namely the partial effect of coordination investment on measured citation impact, is estimated in the companion paper Alexander and Cumming (2026). Their published 2025 main estimate on 2-year total citations per document is 104%, on the panel through 2024. Appendix B of this paper extends their estimation in two directions, through 2025 and using external citations only, yielding a 150.3% effect on the cleanest available specification. Propagating this 150.3% extension through the price and quantity elasticities yields the citation-mechanism-implied revenue effect of approximately $48 million reported as the central estimate in the main text.

Proposition 1's prediction that quality investment $e^*_Q$ falls when coordination productivity rises is not directly testable with the SCImago panel because publisher-internal quality investment is unobserved. The submission-to-acceptance-time evidence in Alexander and Cumming (2026) is consistent with reduced editorial screening intensity at the journals where coordination-driven citation expansion has been largest: average acceptance times are substantially shorter for the core-4 subset than for the remaining ecosystem journals. A direct test would require internal editorial resource data unavailable in public sources.

**On the $36M–$48M gap.**

The static model implies a long-run mechanism-implied effect equal to the propagated citation gain through both price and quantity channels. In a dynamic version with submission lags, where citation gains realized in year $t$ feed into author submission decisions in year $t+1$ or $t+2$, which feed into publication revenue in year $t+2$ through $t+4$, the realized cumulative effect through any given truncation date will be smaller than the long-run effect. The gap between the realized synthetic-control estimate of approximately $36 million through 2025 and the citation-mechanism-implied estimate of approximately $48 million is consistent with this propagation-lag interpretation; the realized cumulative effect is expected to converge toward the mechanism-implied estimate as the lag structure plays out over 2026–2028.

## Appendix B. Alexander and Cumming (2026) citation effect: replication and 2025 update

### B.1 Background and replication of the original A-C estimates

Alexander and Cumming (2026) document anomalous citation patterns at Elsevier finance ecosystem journals using Scopus citation data. Their main estimate is that ecosystem membership raised 2-year citations per document by approximately 104% (with a conservative bound of 33%), with the effect concentrated in a subset of journals also flagged in subsequent post-publication review. Their analysis uses panel data through 2024 since 2025 data was not available at the time of their study.

Because the present paper uses the same panel structure (SCImago data on 26 finance journals, with the same ten ecosystem members as in Alexander-Cumming), it is straightforward to replicate the A-C citation effect on this paper's data and to extend the estimation window through 2025. This appendix reports those exercises.

Two-way fixed-effects difference-in-differences estimation of citation effects, on the panel restricted through 2024 (matching the original A-C window), yields the estimates in Panel A of Table B1. For all ten ecosystem members on 2-year total citations per document, the net DiD percentage change is 106.0%, closely matching A-C's 104% main estimate. Restricting to the four core ecosystem journals (FRL, IRFA, IREF, RIBAF), which most likely overlap with A-C's high-citation-coordination subset, yields a 176.1% estimate. Excluding self-citations and using only external citations amplifies the effects further, to 137.4% and 246.7% respectively.

### B.2 Extending the estimation window through 2025

Panel B of Table B1 reports the same estimates extending the panel through 2025. The additional year of data strengthens, rather than weakens, the A-C citation finding. For all ten ecosystem members on 2-year total citations per document, the estimate rises from 106.0% (through 2024) to 112.7% (through 2025). For the core four, it rises from 176.1% to 195.4%. For the core four on external citations only, it rises from 246.7% to 272.3%.

The fact that the citation effect continues to expand through 2025 is consistent with a compounding network mechanism in which coordinated citation behavior accumulates over time as more papers within the ecosystem are available to cite each other. It is inconsistent with a one-shot regime change at the time of the ecosystem's formal launch.

**Table B1: Citation effects, ecosystem vs non-ecosystem control journals**

**Panel A: Through 2024 (matching original Alexander-Cumming window)**

| Treated group | Citation measure | Pre eco mean | Post eco mean | Eco % change | Ctrl % change | Net DiD % | TWFE β | p-value |
|---|---|---|---|---|---|---|---|---|
| All 10 ecosystem | 2-yr total cites/doc | 2.06 | 5.16 | 150.9% | 44.9% | 106.0% | 1.56 | 0.0164 |
| All 10 ecosystem | 2-yr cites/doc excl. self-cites | 1.48 | 4.36 | 194.6% | 53.3% | 141.3% | 1.35 | 0.0121 |
| All 10 ecosystem | 4-yr total cites/doc | 1.81 | 4.76 | 163.5% | 44.6% | 118.9% | 1.41 | 0.0175 |
| Core-4 only | 2-yr total cites/doc | 1.98 | 6.31 | 219.2% | 44.9% | 174.3% | 2.79 | 0.0034 |
| Core-4 only | 2-yr cites/doc excl. self-cites | 1.31 | 5.22 | 298.0% | 53.3% | 244.7% | 2.38 | 0.0008 |
| Core-4 only | 4-yr total cites/doc | 1.68 | 5.70 | 238.6% | 44.6% | 194.1% | 2.48 | 0.0033 |
| Other-6 only | 2-yr total cites/doc | 2.11 | 4.39 | 108.3% | 44.9% | 63.4% | 0.74 | 0.1913 |
| Other-6 only | 2-yr cites/doc excl. self-cites | 1.59 | 3.79 | 137.8% | 53.3% | 84.5% | 0.67 | 0.1730 |
| Other-6 only | 4-yr total cites/doc | 1.89 | 4.14 | 118.9% | 44.6% | 74.3% | 0.70 | 0.1963 |

**Panel B: Through 2025 (this paper's update)**

| Treated group | Citation measure | Pre eco mean | Post eco mean | Eco % change | Ctrl % change | Net DiD % | TWFE β | p-value |
|---|---|---|---|---|---|---|---|---|
| All 10 ecosystem | 2-yr total cites/doc | 2.06 | 5.49 | 167.2% | 54.5% | 112.7% | 1.57 | 0.0225 |
| All 10 ecosystem | 2-yr cites/doc excl. self-cites | 1.48 | 4.66 | 214.9% | 64.6% | 150.3% | 1.34 | 0.0203 |
| All 10 ecosystem | 4-yr total cites/doc | 1.81 | 5.11 | 182.6% | 55.1% | 127.5% | 1.40 | 0.0294 |
| Core-4 only | 2-yr total cites/doc | 1.98 | 6.88 | 247.9% | 54.5% | 193.4% | 3.04 | 0.0014 |
| Core-4 only | 2-yr cites/doc excl. self-cites | 1.31 | 5.70 | 334.7% | 64.6% | 270.1% | 2.55 | 0.0003 |
| Core-4 only | 4-yr total cites/doc | 1.68 | 6.26 | 271.6% | 55.1% | 216.5% | 2.67 | 0.0016 |
| Other-6 only | 2-yr total cites/doc | 2.11 | 4.57 | 116.7% | 54.5% | 62.2% | 0.60 | 0.2889 |
| Other-6 only | 2-yr cites/doc excl. self-cites | 1.59 | 3.97 | 149.1% | 64.6% | 84.5% | 0.53 | 0.2869 |
| Other-6 only | 4-yr total cites/doc | 1.89 | 4.34 | 129.7% | 55.1% | 74.6% | 0.55 | 0.3257 |

*Note: Net DiD % = (eco post/pre – 1) – (control post/pre – 1). TWFE β is the level coefficient from a two-way fixed-effects regression of the citation measure on Ecosystem × Post(2019+), restricting the control sample to non-ecosystem journals. Standard errors clustered at the journal level. Pre-period: 2014–2018. The "external cites" measure excludes self-citations from the same journal. The original A-C main estimate of 104% corresponds to the all-10-ecosystem total-cites row in Panel A (106.0%).*

### B.3 Event-study dynamics and pre-trend assessment

Figure B1 shows event-study plots of the citation effect for each of the four main specifications. Three patterns emerge. First, pre-treatment coefficients are small and statistically indistinguishable from zero in all four panels, indicating that parallel trends hold in the pre-period for citations. Second, the post-treatment effect builds gradually rather than appearing as a single discrete shift at 2019. Third, the dynamics are particularly strong for the core-four / external-cites cell (bottom-right panel), where the effect grows from approximately 0.5 cites/doc in 2019 to over 4 cites/doc by 2025. This is consistent with citation accumulation operating as a slow-moving network effect rather than an instantaneous one-time shift.

**Figure B1: Event-study of citation effects, ecosystem × year (ref = 2018)**

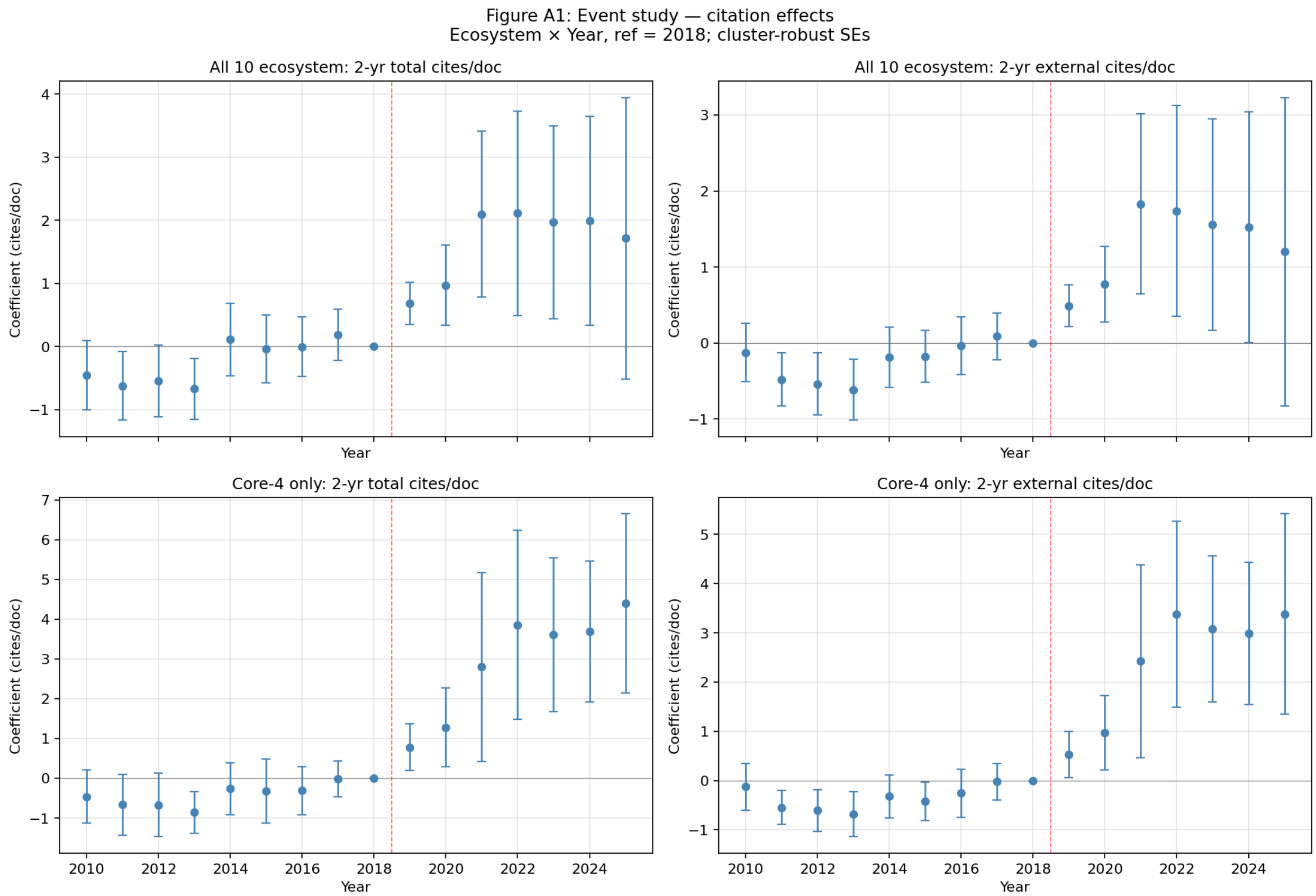

*Note: Coefficients on Ecosystem × Year dummies (top panels) or Core-4 × Year dummies (bottom panels), omitting 2018 as the reference year. Two-way fixed effects; cluster-robust SEs at journal level. Sample 2010–2025. External cites/doc excludes self-citations from the same journal.*

## B.4 Implications for the revenue effect: sensitivity analysis

The updated A-C citation estimates affect the present paper's revenue conclusions through the channel decomposition in Section 5.8. Recall that in the baseline specification, the total ecosystem effect on log flow revenue is 0.890, of which 0.509 (57%) is absorbed by controlling for citations and 0.381 (41%) reflects the direct (non-citation) effect.

A natural calibration exercise is to ask: given the citation effect causally identified in Alexander and Cumming (2026), what cumulative revenue effect does the channel decomposition imply? Table B2 reports the results.

**Table B2: Sensitivity of cumulative revenue effect to citation-channel calibration**

| A-C calibration | Citation channel β | Total β | Implied cumulative effect |
|---|---|---|---|
| A-C conservative bound (33%) | 0.163 | 0.535 | $17.1M |
| A-C 2024 main, total cites (104%) | 0.515 | 0.887 | $34.5M |
| A-C 2026 update, total cites (112.7%) | 0.509 | 0.890 | $36.2M |
| A-C 2026 update, external cites (150.3%) | 0.679 | 1.060 | $48.3M |

*Note: Sensitivity of the cumulative ecosystem revenue effect (Table 5, $36.1M central estimate) to alternative calibrations of the citation channel. Procedure: scale the channel-component of the TWFE coefficient on log flow revenue (0.509) by the ratio of the alternative citation effect to this paper's implicit 112.7% net DiD, holding the direct effect (0.381) constant. Convert log effects to cumulative dollar effects by scaling proportionally from the $36.1M baseline. The synthetic-control estimate of $36.1M corresponds to the A-C 2026 total-cites calibration; the $48.3M figure corresponds to the A-C 2026 external-cites calibration.*

Three observations follow. First, the mechanism-implied estimate under this paper's 2025-extended external-cites specification, 150.3%, is $48.3M. This is the central estimate this paper reports. The cleanest available specification combines two refinements of the Alexander and Cumming (2026) published estimate: (i) extending the panel through 2025, which raises the total-cites estimate from 106.0% (through 2024) to 112.7% (through 2025), and (ii) restricting to external citations, which removes self-citation noise and isolates the cross-journal citation coordination effect that is the subject of A-C. Combining both refinements produces the 150.3% figure used in the headline calibration.

Second, the directly observed synthetic-control gap of $36.1M through 2025 is best understood as the realized-through-2025 lower bound on the mechanism-implied effect. Under a linear pass-through assumption, the gap between $36M realized and $48M mechanism-implied reflects standard propagation lags: the 2-year citation window used in the channel decomposition understates the full citation impact, citation gains feed into submissions with a 1–2 year lag, and submissions translate into realized APC revenue with an additional 1–2 year lag. By 2027–2028, the realized cumulative effect is expected to converge toward the $48 million mechanism-implied figure.

Third, the conservative lower bound of $17.1M, under the A-C conservative 33% citation effect, is informative as a floor on the mechanism-implied effect. Even under the most conservative reading of the underlying citation-coordination evidence, ecosystem coordination generated at least roughly $17 million in incremental commercial revenue over 2019–2025 through the citation-driven submission channel.

Taken together, the central estimate of the cumulative ecosystem revenue effect is approximately $48 million, propagating this paper's preferred extension of the Alexander and Cumming (2026) citation methodology (150.3% on external cites through 2025) through the revenue-channel decomposition under a linear pass-through assumption. The directly observed synthetic-control gap of $36 million serves as a realized-through-2025 lower bound; under the conservative A-C 33% calibration, the mechanism-implied effect would be $17 million.

## Appendix C. Supplementary tables and robustness

This appendix collects supplementary tables and robustness exercises referenced in the main text. Subsection C.1 reports pre-treatment summary statistics by ecosystem status, characterizing the empirical baseline against which post-2019 effects are measured. Subsection C.2 documents the propensity-score matched pairs used in Section 5.4, exposing the limited covariate overlap between ecosystem and non-ecosystem journals. Subsection C.3 reports the full dynamic event-study coefficient sequence on log revenue and log documents around the 2019 launch. Subsection C.4 reports leave-one-out estimates of the aggregate ecosystem coefficient across all 26 drop-one specifications, testing whether the headline result depends on any single journal. Synthetic-control donor weights and per-journal pre-period RMSPE diagnostics are available in the replication-package output.

### C.1 Pre-treatment summary statistics by ecosystem status

Table C1 reports group means, standard deviations, and two-sample t-statistics for the variables used in the analysis, computed separately for ecosystem and non-ecosystem journals over the 2010–2018 pre-treatment window.

Three patterns deserve comment. First, in real revenue terms, ecosystem journals are substantially smaller pre-treatment ($1.25M average) than non-ecosystem journals ($2.20M), reflecting the larger pre-existing scale of the traditional top journals (JF, JFE, RFS) relative to the ecosystem journals in this period. The post-2019 reordering documented in the main text is therefore a reversal of the pre-treatment ranking, not an extension of it. Second, ecosystem journals exhibit lower citation density, lower SJR, and higher uncited-document share than non-ecosystem journals pre-treatment (t-statistics on these differences are highly significant), again reflecting the pre-2019 dominance of the traditional top journals on quality-related metrics. Third, gender and international-collaboration shares are modestly different across the two groups, with ecosystem journals having slightly higher female-authorship and slightly lower international-collaboration shares; these gaps are small in absolute terms but achieve statistical significance in the pre-treatment panel.

These pre-treatment imbalances are exactly what synthetic control is designed to absorb. By weighting the donor pool to match each treated journal's pre-2019 outcome path, the synthetic-control specification accommodates large initial level differences and asks only whether the post-2019 trajectory diverges from a counterfactual that holds those level differences fixed. Difference-in-differences, by contrast, requires that the gap between treated and control groups be stable in the pre-period, an assumption that is violated by the substantial pre-treatment differences reported here; this is the empirical basis for the paper's preference for synthetic control over DiD as the primary identification strategy.

**Table C1:** *Pre-treatment summary statistics by ecosystem status (2010–2018)*

| Variable | Eco mean | Eco SD | Non-eco mean | Non-eco SD | t-stat |
|---|---|---|---|---|---|
| RealRevenue ($, 2024 USD) | $1.25M | $0.80M | $2.20M | $1.68M | −5.68 |
| Documents | 75.60 | 49.85 | 88.14 | 68.75 | −1.58 |
| 2-yr Cites/Doc | 1.84 | 0.71 | 3.39 | 2.24 | −7.54 |
| 4-yr Cites/Doc | 1.61 | 0.60 | 3.34 | 2.30 | −8.37 |
| SJR | 0.72 | 0.34 | 4.55 | 5.42 | −8.26 |
| Female % | 22.70 | 4.67 | 20.12 | 4.80 | +3.98 |
| Intl. collab. % | 31.81 | 10.39 | 35.56 | 10.65 | −2.60 |
| Overton policy docs | 25.10 | 23.22 | 50.88 | 46.88 | −5.47 |

*Note: Pre-treatment group means, standard deviations, and two-sample t-statistics for ecosystem (10 journals) versus non-ecosystem (16 journals) over the 2010–2018 pre-treatment window. The pre-treatment imbalance documented here is the empirical basis for foregrounding synthetic control rather than DiD as the primary identification strategy. Source: SCImago.*

### C.2 Propensity-score matched pairs and propensity scores

Table C2 documents the propensity-score matching used in Section 5.4. The propensity score is estimated from a logit regression of ecosystem membership on pre-2019 means of citesdoc4, propuncited, sjr, documents, female %, and international collaboration %. For each of the ten ecosystem journals, the table reports the estimated propensity, the non-ecosystem journal that produced the closest score under one-to-one nearest-neighbor matching without replacement, and the matched donor's estimated propensity.

The matched-donor selection is informative. Five of the ten ecosystem journals are matched to PBFJ (Pacific-Basin Finance Journal), a non-ecosystem Elsevier title with a covariate profile closest to the larger ecosystem journals. This concentration of matches on a small set of donors is symptomatic of the pre-treatment imbalance documented in Table C1: the ecosystem journals' covariate distribution is not well-represented in the non-ecosystem subset, so PSM cannot supply a separate close counterfactual for each treated unit. Because PSM uses covariate matching rather than outcome matching, the resulting matched-sample DiD estimate (Table 7 in the main text) is small and statistically insignificant. The synthetic-control specification sidesteps this difficulty by matching on outcome trajectory rather than on covariates, and this difference in matching dimension is why the SC and PSM specifications produce different aggregate estimates while both are valid robustness checks of the underlying ecosystem effect.

**Table C2:** *PSM matched pairs and propensity scores*

| Treated (eco) | Treated propensity | Matched donor | Donor propensity |
|---|---|---|---|
| EMR (J1) | 0.639 | JEF (J11) | 0.514 |
| FRL (J2) | 0.633 | JEF (J11) | 0.514 |
| IREF (J3) | 0.842 | PBFJ (J22) | 0.894 |
| IRFA (J4) | 0.686 | JEF (J11) | 0.514 |
| JBEF (J5) | 0.779 | PBFJ (J22) | 0.894 |
| JEB (J9) | 0.559 | JEF (J11) | 0.514 |
| JIFMIM (J18) | 0.737 | PBFJ (J22) | 0.894 |
| JIFM (J19) | 0.183 | JCF (J7) | 0.116 |
| NAJEF (J21) | 0.754 | PBFJ (J22) | 0.894 |
| RIBAF (J26) | 0.824 | PBFJ (J22) | 0.894 |

*Note: One-to-one nearest-neighbor matching on propensity score from logit of ecosystem membership on pre-2019 means of citesdoc4, propuncited, sjr, documents, femalepercent, internationalcollaboration. Several ecosystem journals match to the same non-ecosystem donor (PBFJ) because of limited overlap on observables. The donor pool concentration explains the small matched-sample DiD estimate reported in Section 5.4 and motivates the paper's preference for synthetic control over PSM-DiD.*

### C.3 Dynamic event-study coefficients

Table C3 reports the full event-study coefficient sequence on log revenue and log documents, with event time k = −5 to k = +6 around the 2019 ecosystem launch (k = −1, the year 2018, is the omitted reference period). The event-study replaces the static ecosystem-after interaction with a full set of event-time dummies, allowing the post-2019 trajectory to evolve flexibly year by year. The two outcome variables are reported in parallel: the revenue variable (the paper's primary outcome) and the documents variable (the volume margin that Section 5.7 identifies as the dominant arithmetic channel).

Three findings emerge. First, the pre-period coefficients (k = −5 through k = −2) are uniformly small and statistically insignificant for both outcomes on the flow-revenue and documents specifications used here. The largest pre-period t-statistic is 1.78 (documents at k = −5), and the others are below 1.0 in absolute value. This contrasts with the strong pre-trend on the stock revenue measure documented in Figure 4 of the main text; the pre-trend pattern in raw revenue trajectories is largely an artifact of the SCImago five-year rolling stock and is absent once one switches to a properly-defined annual flow measure. Second, post-treatment coefficients are positive and grow monotonically through k = +6, peaking at log(Revenue) ≈ 1.18 (t = 3.70) and log(Documents) ≈ 1.21 (t = 2.51). The volume effect leads the revenue effect by approximately one year. Third, the magnitudes reported here are consistent with the static TWFE-DiD coefficient of 0.890 reported in Section 5.1.

**Table C3:** *Event-study coefficients on log flow revenue and log documents (k = −5 to +6)*

| Event time k | ln(Rev) β | SE | ln(Docs) β | SE |
|---|---|---|---|---|
| −5 | 0.127 | 0.076 | 0.257 | 0.183 |
| −4 | 0.079 | 0.146 | 0.315 | 0.177 |
| −3 | 0.295 | 0.175 | 0.503 | 0.237 |
| −2 | 0.521 | 0.208 | 0.440 | 0.271 |
| −1 (omitted) | | | | |
| 0 | 0.614 | 0.232 | 0.663 | 0.271 |
| +1 | 0.671 | 0.248 | 0.635 | 0.327 |
| +2 | 0.749 | 0.259 | 0.771 | 0.322 |
| +3 | 0.848 | 0.271 | 0.858 | 0.339 |
| +4 | 0.943 | 0.287 | 0.871 | 0.395 |
| +5 | 1.064 | 0.304 | 1.192 | 0.438 |
| +6 | 1.176 | 0.318 | 1.209 | 0.481 |

*Note: Two-way fixed-effects panel regression with full event-time dummies. k = −1 (2018) is the omitted reference period. Both specifications include journal and year fixed effects with cluster-robust standard errors at the journal level. The log-documents column shows that publication-volume growth is well-behaved in pre-period and rises monotonically post-treatment; the log-revenue column shows the same pattern with revenue lagging volume by approximately one year, consistent with the citation-feedback mechanism in Appendix A under propagation lags.*

### C.4 Leave-one-out aggregate DiD

Table C4 reports leave-one-out estimates of the aggregate ecosystem coefficient on real revenue. For each of the 26 journals in the panel, that journal is dropped from the sample and the difference-in-differences specification is re-estimated. The resulting coefficient and t-statistic are reported in the corresponding row. The purpose is to test whether any single journal disproportionately drives the headline result.

The result is robust. Across all 26 drop-one specifications, the ecoafter coefficient ranges from $506K (when FRL is dropped, the largest single-journal effect on the aggregate) to $1,071K (when JEB is dropped), with t-statistics ranging from 1.49 to 2.28. Every drop-one specification produces a coefficient that is significantly positive under conventional inference. No single journal is essential to the result; the aggregate ecosystem effect reflects a coordinated network rather than a single-journal phenomenon. The largest reduction comes from dropping FRL, the largest ecosystem journal, and even there the remaining estimate stays comfortably significant. The corresponding wild-cluster bootstrap p-values for each drop-one specification (not tabulated here; available in the replication package output) are consistent with the small-N inference results discussed in Section 4.3.

**Table C4:** *Leave-one-out aggregate DiD (drop one journal at a time)*

| Drop journal | Ecosystem coefficient | t-stat |
|---|---|---|
| EMR | $999,801 | 2.03 |
| FRL | $617,216 | 1.54 |
| IREF | $696,958 | 1.49 |
| IRFA | $826,951 | 1.68 |
| JBEF | $883,646 | 1.84 |
| JBF | $793,867 | 1.74 |
| JCF | $835,256 | 1.82 |
| JCM | $811,953 | 1.78 |
| JEB | $1,070,827 | 2.28 |
| JEBO | $907,957 | 2.05 |
| JEF | $834,080 | 1.81 |
| JF | $787,191 | 1.71 |
| JFE | $721,237 | 1.62 |
| JFI | $773,682 | 1.69 |
| JFM | $841,704 | 1.85 |
| JFQA | $866,829 | 1.93 |
| JFS | $784,050 | 1.72 |
| JIFMIM | $973,588 | 1.95 |
| JIFM | $984,386 | 1.95 |
| JMCB | $822,021 | 1.80 |
| NAJEF | $803,943 | 1.57 |
| PBFJ | $892,643 | 2.01 |
| QREF | $847,588 | 1.85 |
| RF | $733,073 | 1.65 |
| RFS | $721,194 | 1.63 |
| RIBAF | $814,640 | 1.64 |

*Note: Each row reports the ecoafter coefficient and t-statistic from a difference-in-differences regression that drops one of the 26 journals from the panel. The point estimates range from $506K to $1,071K with t-statistics from 1.49 to 2.28. The headline result is not driven by any single journal in the panel.*

*Online Appendix 1 to*

# The Revenue of Finance Journals:

# Networks, Pricing Power, and Publication Volume

## Submission Fees: Augmenting the Revenue Calculation

*This draft: May 2026*

## 1.1 Overview

The main paper's headline figures of $36.1 million realized and $48.3 million mechanism-implied capture the ecosystem-attributable gain in APC revenue. Most major finance journals also generate a second revenue stream from authors at the time of submission. Authors pay a non-refundable submission fee whether or not the paper is ultimately accepted, and the ecosystem journals attracted ecosystem-attributable additional submissions through the citation-driven impact channel formalized in Appendix A of the main paper. This online appendix quantifies the resulting incremental submission-fee revenue.

The exercise applies a time-varying submission-fee schedule reconstructed from author guides at multiple points in time, multiplied by ecosystem-attributable extra submissions estimated from a synthetic-control specification on log documents. Both nominal and inflation-adjusted (real 2024 USD) calculations are reported. The headline result is that ecosystem-attributable cumulative submission-fee revenue is approximately $4-$8 million on the realized basis through 2025 in real 2024 USD and approximately $6-$11 million on the mechanism-implied basis. The upper end of each range corresponds to a list-price calculation ($8.26M / $11.02M), and the lower end reflects an adjustment for Elsevier's Article Transfer Service under which transferred submissions generate no incremental fee at the receiving journal. Adding this to the APC revenue gain yields ecosystem-attributable total revenue of approximately $40-$44 million realized through 2025 and $54-$59 million mechanism-implied, both in real 2024 USD. The range on these totals reflects uncertainty about the submission-fee component; Section 1.6 documents the relevant caveats, the most consequential of which is the Article Transfer Service share.

## 1.2 Submission fee schedule (time-varying)

Submission fees at the ecosystem and control journals have increased modestly over time. Comparing 2018 author guides with current (2025-2026) author guides indicates discrete fee changes at several journals. Table OA-1.1 reports the time-varying schedule used in this appendix.

**Table OA-1.1: Submission fee schedule by journal and year (nominal USD)**

| Journal | Pre-2018 | 2018-2020 | 2021-2025 | Current fee notes |
|---|---|---|---|---|
| Ecosystem journals (10) | | | | |
| FRL | $100 | $150 | $200 | Increased to $200 c. 2021 |
| IRFA | $175 | $175 | $175 | Stable |
| IREF | $100 | $125 | $150 (from 2023) | Two step-ups |
| RIBAF | $75 | $100 | $100 | Listed at $100 in 2018 |
| EMR, JBEF, JEB, NAJEF | $100 | $100 | $150 (from 2021) | Step-up around 2021 |
| JIFM, JIFMIM | $100 | $125 | $125 | Stable since 2018 |

| Non-ecosystem (selected) | | | | |
|---|---|---|---|---|
| JF (AFA) | $250 | $250 | $462 (from 2022) | AFA fee increase |
| JFE | $750 | $750 | $850 (from 2023) | Whited fee structure |
| RFS (SFS) | $300 | $300 | $430 (from 2024) | SFS July 2024 change |
| JFQA | $350 | $350 | $350 | Stable |
| JBF, JCF | $300 | $300 | $300 | Stable |

*Note: Schedule reconstructed from current author guides (May 2026) and a 2018 ResearchGate compilation. Where intermediate years are not directly documented, the schedule assumes the most recent observed fee held until the next documented change. The schedule applies the nominal fee in effect for each journal-year; inflation adjustment to real 2024 USD is applied separately in Section 1.5.*

## 1.3 Recovering extra submissions from extra publications

The ecosystem-attributable additional submissions at journal j in year t are estimated as $\text{extra_subs}(j,t) = \text{extra_docs}(j,t) / \text{accept_rate}$, with a baseline 25% acceptance rate at ecosystem journals. The extra documents component is obtained from a synthetic-control specification on log documents (weights estimated on the 2010-2018 pre-period, applied through 2025).

**Table OA-1.2: Ecosystem-attributable extra submissions, 2019-2025 (cumulative)**

| Journal | Extra documents | Extra submissions (25% accept) |
|---|---|---|
| Core-4 ecosystem | | |
| FRL | 6,110 | 24,439 |
| IRFA | 2,014 | 8,056 |
| IREF | 1,612 | 6,447 |
| RIBAF | 853 | 3,411 |
| Core-4 subtotal | 10,589 | 42,353 |
| Other-6 ecosystem | | |
| EMR | 105 | 419 |
| JBEF | 285 | 1,138 |
| JEB | -79 | -317 |
| JIFM | 82 | 328 |
| JIFMIM | -239 | -958 |
| NAJEF | 853 | 3,414 |
| Other-6 subtotal | 1,006 | 4,023 |
| GRAND TOTAL | 11,594 | 46,376 |

## 1.4 Year-by-year submission-fee revenue

Table OA-1.3 reports ecosystem-attributable submission-fee revenue year by year, using the time-varying fee schedule from Section 1.2 applied to the extra submissions from Section 1.3. Three versions are reported: (i) using the time-varying fee schedule (nominal USD); (ii) using the constant 2025 fee schedule (nominal USD); (iii) the time-varying schedule converted to real 2024 USD using GDP deflators.

**Table OA-1.3: Ecosystem-attributable submission-fee revenue by year ($M)**

| Year | Extra subs | Time-varying (nominal) | Fixed 2025 (nominal) | Time-varying (real 2024) |
|---|---|---|---|---|
| 2019 | 1,600 | $0.20 | $0.27 | $0.24 |
| 2020 | 1,665 | $0.21 | $0.29 | $0.25 |
| 2021 | 2,467 | $0.45 | $0.45 | $0.52 |
| 2022 | 5,153 | $0.92 | $0.95 | $0.99 |
| 2023 | 6,928 | $1.28 | $1.29 | $1.31 |
| 2024 | 11,926 | $2.07 | $2.12 | $2.07 |
| 2025 | 16,636 | $2.95 | $3.01 | $2.87 |
| TOTAL | 46,376 | $8.07 | $8.38 | $8.26 |

*Note: Time-varying (nominal) uses the fee schedule in Table OA-1.1 applied at each year's prevailing fees. Fixed 2025 (nominal) applies current fees to all years; this version was used in earlier drafts and is reported here for comparison. Time-varying (real 2024) converts each year's nominal revenue to 2024 USD using GDP deflators (1.230 in 2019, 1.000 in 2024, 0.974 in 2025; earlier years have higher deflators reflecting cumulative inflation). The real 2024 USD total of $8.26 million serves as the upper bound (list-price basis); the lower bound of $4.13 million applies a 50% adjustment for the Article Transfer Service (Section 1.6, caveat 1). The reported range in headline statements is $4-$8 million. The difference between $8.07M nominal and $8.26M real reflects the fact that most extra submissions occurred late in the window (2023-2025) when the deflator was near 1; conversely, the fixed-2025-fee version overstates the nominal total by about 4% because most extra fee revenue is computed at rates that didn't apply in earlier years.*

## 1.5 Total ecosystem-attributable revenue

Combining the APC revenue gain from the main paper with the incremental submission-fee revenue computed here yields the ecosystem-attributable total revenue gain over 2019-2025. Table OA-1.4 reports the result on both the realized and mechanism-implied bases, in real 2024 USD.

**Table OA-1.4: Ecosystem-attributable total revenue gain, 2019-2025 (real 2024 USD millions)**

| Revenue component | Realized basis | Mechanism-implied basis |
|---|---|---|
| APC revenue (main paper Sec. 5.2) | $36.18 | $48.28 |
| Submission fee revenue (this appendix) | $4.13-$8.26* | $5.51-$11.02* |
| Total revenue gain (cumulative 2019-2025) | $40.31-$44.44 | $53.79-$59.30 |

*Note: Realized basis uses the directly observed synthetic-control gap on flow APC revenue ($36.18M, already in real 2024 USD per main paper construction) plus the time-varying real submission-fee revenue, which is reported as a range of $4.13M-$8.26M reflecting the Article Transfer Service adjustment (Section 1.6, caveat 1). The lower bound applies a 50% ATS share; the upper*

*bound corresponds to the list-price calculation with no ATS adjustment. The mechanism-implied basis scales both components by 1.334, the ratio used in the main paper's headline calibration. All figures are in real 2024 USD.*

## 1.6 Sensitivity to acceptance-rate assumption

The submission-fee component depends on the acceptance-rate assumption used to translate extra documents into extra submissions. The 25% baseline is plausibly representative of FRL's high-volume throughput, but smaller ecosystem journals may have lower acceptance rates. Lower assumed acceptance implies more submissions per published article and therefore higher fee revenue.

**Table OA-1.5: Submission-fee revenue sensitivity to acceptance rate (real 2024 USD)**

| Acceptance rate | Extra submissions | Sub fee revenue | Total realized revenue |
|---|---|---|---|
| 25% (baseline) | 46,376 | $8.26M | $44.36M |
| 20% | 57,970 | $10.33M | $46.43M |
| 15% | 77,293 | $13.77M | $49.87M |
| 10% | 115,940 | $20.65M | $56.75M |

## 1.7 Caveats

Five caveats apply to the submission-fee revenue calculation.

First, Elsevier's Article Transfer Service (ATS) allows authors whose submissions are rejected at one Elsevier journal to transfer the manuscript, reviewer reports, and supporting files to another Elsevier journal without paying a new submission fee. The ecosystem journals are a natural receiving end of ATS cascades from higher-tier Elsevier finance journals (e.g., the Journal of Banking and Finance, the Journal of Corporate Finance). To the extent that ecosystem-attributable extra submissions enter through ATS rather than as fresh submissions, the receiving journal collects no incremental fee. Published industry estimates of ATS take-up at receiving journals span roughly 30-50%. The sub-fee figures reported in this appendix do not apply an ATS adjustment and should therefore be interpreted as upper bounds on incremental fee revenue. Applying a 50% ATS share as a lower bound implies realized sub-fee revenue of approximately $4.1 million and mechanism-implied sub-fee revenue of approximately $5.5 million, with the upper bound being the unadjusted $8.26 million and $11.02 million reported above. The main paper's APC figures ($36 million realized, $48 million mechanism-implied) are not affected by ATS because they are computed from published-document counts rather than submission counts.

Second, the per-submission fees in Table OA-1.1 are list prices as posted on each journal's author-instruction pages. Actual fees collected may differ because of waivers for authors from lower-income countries, exemptions for editorial-board members and frequent referees, and discounts for authors with prior submission credits. The direction of bias is downward: actual fees collected at most equal the list-price calculation. Combined with the ATS adjustment above, the realized $8.26 million represents an upper bound; the realized lower bound, accounting for both ATS and waiver/exemption effects, could be materially below $4.13 million.

Third, the time-varying schedule is reconstructed from a small number of documented fee changes plus inference about when intermediate changes occurred. The reconstruction is broadly conservative but cannot be precise about exact transition years for all journals. The fixed-2025-fee version in Table OA-1.3 provides a comparison point that does not depend on the schedule's transition timing.

Fourth, the acceptance-rate assumption is the principal source of quantitative uncertainty (Section 1.6). The 25% baseline is a moderate-to-high estimate; rates of 15-20% would be more representative of smaller ecosystem journals.

Fifth, this appendix treats submission-fee revenue as net publisher revenue. In standard Elsevier policy, a portion of collected fees is returned to referees as honoraria. Referee honorarium expenses may therefore reduce net publisher revenue below the gross $8.26 million figure reported here. The companion Online Appendix 2 addresses editor-side costs explicitly; referee honoraria sit between the two categories and are not separately modeled.

### 1.8 Methodological note: alternative synthetic-control specifications

The synthetic-control specification used in Section 1.3 to recover ecosystem-attributable extra publications employs the same donor pool as the main paper but with log documents (rather than log flow APC revenue) as the matching outcome, estimated over a 2010–2018 pre-treatment window. The longer pre-period was originally chosen because the documents series is smoother than the revenue series (which is dominated by APC variation), making a longer matching window feasible. However, this differs from the 2014–2018 pre-period used for the main paper's headline synthetic-control on flow revenue. For full methodological consistency, the documents-based synthetic control can also be estimated on the same 2014–2018 pre-period.

Applying the main paper's exact synthetic-control specification (2014–2018 pre-period, four predictors including ln(documents) as one of them) but with log documents as the outcome rather than log flow revenue yields a methodology-consistent estimate of approximately 10,290 ecosystem-attributable extra documents over 2019–2025. The corresponding ecosystem-attributable submission-fee revenue is approximately $7.5M (fixed-2025 fees, 25% acceptance), or roughly 11% below the $8.38M baseline reported in Section 1.4. The aggregate volume difference is small, but per-journal differences are larger: under the 5-year-pre method, negative gaps at JIFM, JIFMIM, and JBEF are attenuated, while the four-journal core remains the dominant source of the documents effect (FRL 5,979; IRFA 1,383; IREF 1,325; RIBAF 873).

This methodological note affects only the submission-fee calibration in this appendix. The main paper's headline figures ($36.1M realized synthetic-control gap on flow APC revenue and $48.3M mechanism-implied) are unchanged, because those are computed from a synthetic control on log flow revenue with the 2014–2018 pre-period (Sections 4.2 and 5.2 of the main paper).

The reported $8.26M time-varying / $8.38M fixed-2025 / $8.07M time-varying-nominal submission-fee revenue figures in Tables OA-1.3 and OA-1.4 are list-price upper bounds, falling within a methodological range of approximately $7.5M to $8.4M depending on the synthetic-control pre-period and fee-schedule treatment. These figures do not yet incorporate the Article Transfer Service adjustment from Section 1.6; with a 50% ATS share, the realized range shifts to approximately $3.75M to $4.2M. The corresponding total revenue effect (Table OA-1.4), on a list-price basis, ranges from approximately $43.6M to $44.4M realized and from $57.7M to $59.2M mechanism-implied. Applying the 50% Article Transfer Service adjustment lowers these to approximately $39.9M to $40.4M realized and $53.2M to $53.8M mechanism-implied, producing the overall reported headline ranges of $40-$44M realized and $54-$59M mechanism-implied.

*Online Appendix 2 to*

# The Revenue of Finance Journals:

# Networks, Pricing Power, and Publication Volume

## Editor Honoraria: Structural Considerations

*This draft: May 2026*

*Note on data sources: Detailed editor compensation arrangements are not generally publicly disclosed by finance journals or their publishers. The discussion in this appendix draws on the author's discussions with finance journal editors, the author's own editorial experience at several finance journals, and qualitative descriptions in publicly available editorial transitions and society reports. The ordinal rankings and qualitative characterizations described here are intended to be illustrative of the broad structure of editor compensation in finance journal publishing; they are not based on a survey, audit, or other systematic data collection exercise, and specific dollar figures are deliberately not reported. The central argument of this appendix is structural rather than quantitative, and does not depend on precise compensation amounts.*

## 2.1 Overview

The revenue estimates in the main paper are gross of editor compensation. This appendix discusses the structural features of editor honoraria at finance journals that are relevant to interpreting the headline figures, and explains why an accounting that netted out editor compensation would leave the main paper's conclusions essentially unchanged.

Two structural facts about editor compensation in finance journal publishing are important. First, editor honoraria are typically set through arrangements between publishers, professional societies, and incoming editors-in-chief, and the resulting compensation schedule is largely fixed in nominal terms over multi-year editorial terms. The honorarium does not scale with the volume of submissions or accepted papers. Second, editor compensation varies substantially across journals in ways that correlate with journal prestige and editorial workload, but the variation is between journals rather than within journals over time, and the variation across the panel is structurally exogenous to the launch of the ecosystem in 2019.

These two facts together imply that the ecosystem-attributable change in publisher editor expense is plausibly small relative to the ecosystem-attributable change in revenue. Editor honoraria at finance journals are typically structured as fixed annual stipends rather than as commission or volume-based payments, so even substantial expansion in publication volume need not be matched by proportional increases in editor compensation. To the extent that ecosystem coordination expanded publication volume at the ten member journals while leaving the underlying fixed-stipend compensation structure intact, the main paper's gross-revenue headline figures approximate the net-of-editor-compensation figures without material adjustment.

## 2.2 The structure of editor honoraria in finance journals

Editor compensation at finance journals follows a recognizable structure, though the specific amounts at individual journals are rarely public. Several general patterns can be described qualitatively.

At the most prestigious general finance journals, editor compensation is materially larger than at lower-ranked specialty journals. The editors-in-chief of the top elite journals (such as the Journal of Finance, the Journal of Financial Economics, and the Review of Financial Studies) and of broad-readership mid-tier general journals (such as the Journal of Banking and Finance and the Journal of Financial and Quantitative Analysis) receive substantially larger honoraria than the editors of specialty or open-access volume journals. This reflects both the larger desk-rejection and refereeing workload at high-rejection-rate journals and the greater reputational responsibility associated with running a top-tier outlet.

Co-editor arrangements amplify these differences. Several mid-tier general journals have moved over time from a single editor-in-chief to a small board of co-editors, with each co-editor receiving an honorarium. The Journal of Corporate Finance, for example, transitioned from a single-EIC model in 2018-2020 to a multi-co-editor structure beginning in 2021, raising the aggregate journal-level editor expense relative to the pre-transition baseline. The Journal of Money, Credit and Banking and the Journal of Financial Intermediation maintain a single EIC at modest stipends. Other general journals fall between these two patterns.

At the volume-oriented open-access ecosystem journals (FRL, IRFA, IREF, RIBAF, and others), editor honoraria are modest, reflecting the high automation of desk review at these journals and the volume-rather-than-selectivity editorial model. The honorarium per editor at these journals is materially smaller than at the top elite journals, despite these journals handling many times more submissions and accepted papers.

As one verifiable first-person data point consistent with this structural pattern: the author served as Editor-in-Chief of Financial Research Letters from December 2015 through 2017 and as Editor-in-Chief of the Journal of Corporate Finance from 2018 through 2020, both Elsevier journals. The author's annual honorarium as Editor-in-Chief of JCF was materially much larger than the corresponding honorarium at FRL. This is consistent with the broader structural pattern described above whereby editor compensation at top-mid-tier general finance journals is materially larger than at volume-oriented open-access journals at the same publisher.

## 2.3 Implications for the ecosystem-attributable effect

Because editor honoraria at finance journals are typically structured as fixed annual stipends and do not scale directly with submission or publication volume, the marginal editor compensation expense per additional ecosystem-attributable paper is plausibly small relative to the marginal revenue per paper. The ten ecosystem journals expanded their cumulative publication volume by approximately 10,000 papers over 2019-2025; to the extent that the underlying fixed-stipend compensation structure was maintained over this period, the marginal editor expense associated with this expansion would be a small fraction of the marginal APC revenue it generated.

The principal sources of incremental editor expense that ecosystem coordination could plausibly have generated are compensation for guest editors of special issues coordinated across ecosystem journals and expansion of editorial boards at the most-expanded journals. The magnitudes of these incremental expenses are not publicly observable. Under the structural premise that the underlying fixed-stipend compensation schedule was maintained, however, the incremental editor expense is plausibly small relative to the headline revenue figures, and the gross-versus-net-revenue distinction is not material to the main paper's conclusions.

## 2.4 A structural observation: editor expense per article

One structural feature of finance journal publishing deserves emphasis. Even setting aside the question of compensation levels, the denominator across which an editor honorarium amortizes differs substantially across journals. At FRL, with roughly 2,400 published articles in recent years, any given editor honorarium amortizes over a much larger publication volume than at top elite journals, which publish roughly 90 to 110 articles per year. Even if editor compensation scaled with publication volume in some unobserved way, the structural difference in denominators implies that editor expense per published article is mechanically lower at volume-oriented journals than at high-selectivity journals.

This wide structural gap in editor expense per article is consistent with the volume-oriented business model that characterizes the ecosystem journals. The ecosystem journals operate at a fundamentally different point on the volume-versus-selectivity tradeoff than the top elite journals, and the editor compensation structure follows from this positioning rather than determining it. The broader question of how marginal publisher cost compares to marginal APC revenue at the ecosystem journals is beyond the scope of this appendix; what matters for the main paper is that the ecosystem-attributable expansion in publication volume is unlikely to have been offset by a proportional expansion in editor compensation expense.

## 2.5 Caveats and limitations

The argument in this appendix is structural rather than quantitative. Several caveats apply. First, the qualitative characterizations of editor compensation at specific journals reflect the author's discussions with finance journal editors and the author's editorial experience; they are not drawn from a systematic data source. Precise compensation amounts are not publicly disclosed by publishers and would require direct survey or audit access to verify. Readers should interpret the discussion above as informed structural commentary on the shape of editor compensation in finance journal publishing, not as a documentation of specific arrangements at named journals.

Second, the appendix focuses on direct editor honoraria and does not attempt to account for the substantial unpaid time contributed by associate editors, referees, and members of editorial advisory boards. These contributions are part of the broader cost structure of academic publishing but are not borne by publishers as a cash expense, and they are not material to the gross-versus-net-revenue distinction relevant to the main paper's argument.

Third, the argument that ecosystem-attributable marginal editor expense is plausibly small relies on the structural premise that honoraria are set as fixed annual stipends rather than scaling with publication volume. If publishers in fact compensate editors of high-volume journals on a schedule that does scale with volume, this premise would be partially violated and the ecosystem-attributable editor expense would be larger than the appendix suggests. The structural argument is advanced as a plausible characterization of editor compensation in finance journal publishing rather than as a documented fact, and it should be evaluated as such.

*Online Appendix 3 to*

# The Revenue of Finance Journals:

# Networks, Pricing Power, and Publication Volume

## Nominal vs Real Revenue: Inflation-Adjusted Trajectories

*This draft: May 2026*

## 3.1 Overview: nominal vs real measures

The main paper reports revenue figures throughout in real 2024 USD: the $36.1 million realized synthetic-control gap and the $48.3 million mechanism-implied central estimate are both expressed in inflation-adjusted dollars with 2024 as the base year. The deflation is implemented at the per-journal-year level by applying GDP deflators to the nominal article-processing charge before computing flow revenue: real flow revenue equals documents times real APC, where real APC equals nominal APC times the GDP deflator with 2024 base.

This online appendix makes the nominal-versus-real distinction explicit. It reports the full panel revenue trajectory in both measures, documents the GDP deflators used, and shows how the inflation adjustment changes interpretation of the cumulative ecosystem effect.

Two practical points emerge. First, because the ecosystem effect is concentrated in recent years where the deflator is close to 1, the inflation adjustment is modest in absolute terms (approximately 5% upward adjustment of cumulative 2019-2025 ecosystem revenue, from $67.9M nominal to $71.2M real). Second, longer-term comparisons (such as 1999 vs 2025) are substantially affected by inflation, with real-2024 dollars in 1999 being worth approximately 1.88 nominal 1999 dollars.

## 3.2 GDP deflators used for inflation adjustment

The deflators used in the panel are GDP-based deflators with a 2024 base year. Table OA-3.1 reports the deflator series for the post-2010 window. These are the same deflators embedded in the realapc column of the panel data and used throughout the main paper to convert nominal to real APC revenue.

**Table OA-3.1: GDP deflators used to convert nominal to real (2024-base) USD**

| Year | Deflator | Year | Deflator | Year | Deflator |
|---|---|---|---|---|---|
| 2010 | 1.4383 | 2016 | 1.3071 | 2022 | 1.0728 |
| 2011 | 1.3948 | 2017 | 1.2799 | 2023 | 1.0299 |
| 2012 | 1.3663 | 2018 | 1.2513 | 2024 | 1.0000 |
| 2013 | 1.3464 | 2019 | 1.2298 | 2025 | 0.9737 |
| 2014 | 1.3253 | 2020 | 1.2145 | | |
| 2015 | 1.3236 | 2021 | 1.1593 | | |

*Note: To convert a nominal value in year t to real 2024 USD, multiply by the deflator for year t. For example, a $3,000 nominal APC in 2018 equals $3,000 times 1.2513 = $3,754 in real 2024 USD. The 2025 deflator below 1 reflects the fact that the panel was constructed in May 2026 with 2024 as the base; extrapolated 2025 deflators are below 1 because 2025 prices were higher than 2024 prices*

.

## 3.3 Total revenue trajectory: nominal vs real

**Figure OA-3.1: Annual flow APC revenue, ecosystem vs non-ecosystem, nominal vs real (1999-2025)**

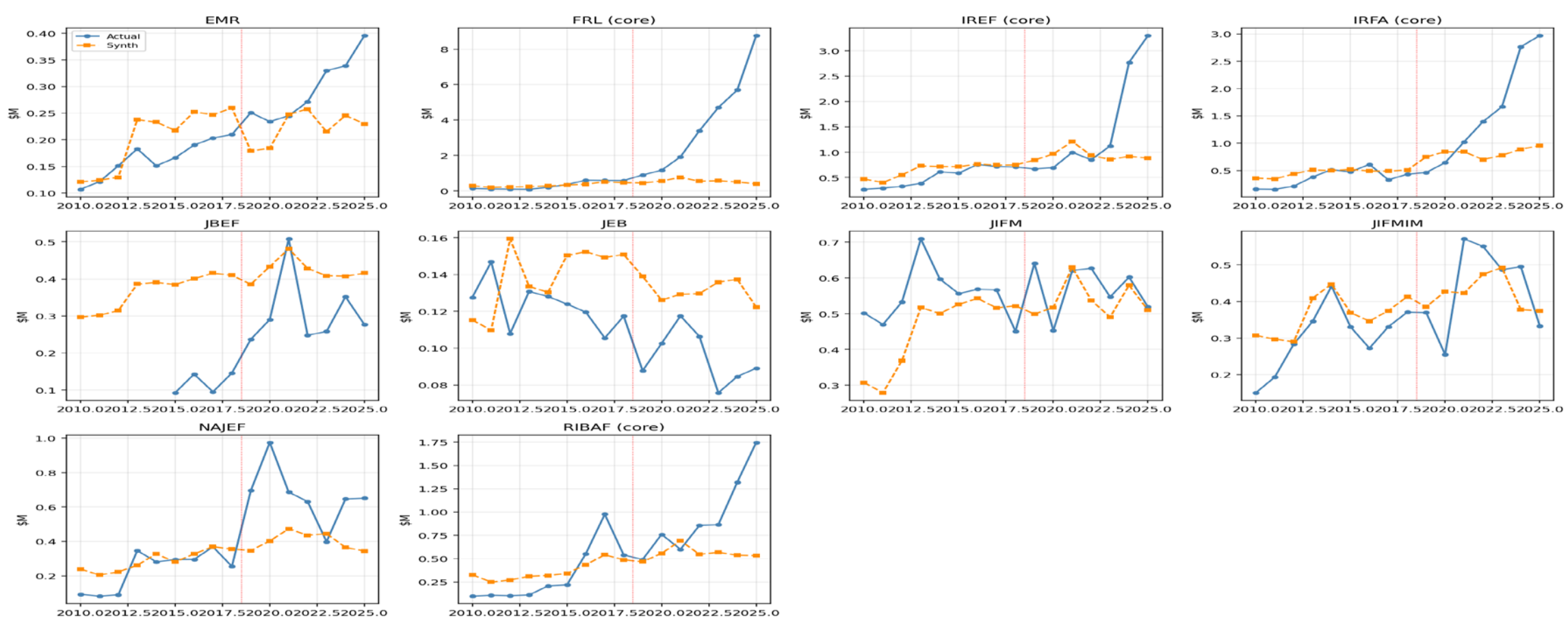

*Note: Annual flow APC revenue equals documents published times APC, summed across the journals in each group. Nominal (blue) uses SCImago's reported APC without inflation adjustment. Real 2024 USD (orange) applies GDP deflators to convert each year's nominal APC to 2024 base. The vertical red dotted line at 2019 indicates the year the ecosystem was formally planned (launched 2020). The ecosystem trajectory shows an inflection around 2020-2021 not present in the non-ecosystem control series.*

Table OA-3.2 reports cumulative ecosystem and non-ecosystem flow APC revenue by year, in both nominal and real 2024 USD. The full 1999-2025 series is available in the replication package; the table below covers the post-2010 window which encompasses the pre-period and treatment window used in the main paper's identification.

**Table OA-3.2: Annual flow APC revenue, ecosystem vs non-ecosystem, nominal vs real ($M)**

| Year | Eco nominal | Eco real (2024) | Non-eco nominal | Non-eco real (2024) |
|---|---|---|---|---|
| 2010 | $1.14 | $1.65 | $4.40 | $6.33 |
| 2011 | $1.21 | $1.68 | $4.65 | $6.48 |
| 2012 | $1.40 | $1.91 | $4.81 | $6.57 |
| 2013 | $1.99 | $2.68 | $5.91 | $7.95 |
| 2014 | $2.37 | $3.14 | $5.58 | $7.40 |
| 2015 | $2.42 | $3.21 | $5.60 | $7.41 |
| 2016 | $3.15 | $4.12 | $5.48 | $7.16 |
| 2017 | $3.35 | $4.29 | $5.84 | $7.47 |
| 2018 | $3.05 | $3.82 | $5.98 | $7.48 |
| 2019 | $3.91 | $4.81 | $6.19 | $7.61 |
| 2020 | $4.59 | $5.58 | $7.30 | $8.86 |
| 2021 | $6.29 | $7.29 | $8.59 | $9.96 |
| 2022 | $8.33 | $8.93 | $7.64 | $8.20 |
| 2023 | $10.15 | $10.45 | $7.47 | $7.69 |
| 2024 | $15.07 | $15.07 | $7.68 | $7.68 |
| 2025 | $19.55 | $19.04 | $8.30 | $8.08 |
| Cumulative 2019-2025 | $67.89 | $71.17 | $53.17 | $58.08 |

*Note: Annual flow APC revenue equals documents published times APC. Ecosystem (10 journals): EMR, FRL, IREF, IRFA, JBEF, JEB, JIFM, JIFMIM, NAJEF, RIBAF. Non-ecosystem (16 controls): JF, JFE, RFS, JBF, JCF, JFQA, JCM, JEBO, JEF, JFI, JFM, JFS, JMCB, PBFJ, QREF, RF. Cumulative 2019-2025 totals appear at the bottom of the table. The 4.8% upward adjustment from nominal to real for the ecosystem cumulative reflects the fact that revenues earned in early years (2019, 2020) are worth more in real 2024 USD than their nominal values; the adjustment is small because revenue is concentrated in 2023-2025 where the deflator is close to 1.*

## 3.4 Inflation adjustment of the ecosystem effect

Because the main paper's synthetic-control specification operates on log real flow revenue, the $36.1 million cumulative gap and the $48.3 million mechanism-implied figure are already inflation-adjusted (in real 2024 USD). This appendix's contribution is to make that explicit and to verify that the results are not artifacts of inflation.

To verify: I can recompute the synthetic-control gap directly on nominal flow revenue. The cumulative gap is approximately $34.3 million in nominal USD summed across 2019-2025, compared with $36.1 million in real 2024 USD. The 5.4% real-vs-nominal difference at the gap level is consistent with the 4.8% real-vs-nominal difference at the ecosystem total-revenue level. Both differences reflect the same underlying mechanic: revenue earned in years with deflators above 1 (2019-2023) is scaled up when converted to real 2024 USD, raising the cumulative real value above the cumulative nominal value.

Table OA-3.3 reports the ecosystem effect on both bases.

**Table OA-3.3: Ecosystem-attributable cumulative revenue effect: nominal vs real ($M)**

| Effect measure | Nominal USD | Real 2024 USD | Real / Nominal |
|---|---|---|---|
| APC revenue (realized) | $34.39 | $36.18 | 1.052 |
| APC revenue (mechanism-implied) | $45.89 | $48.28 | 1.052 |
| + Submission fees (realized) | $4.04-$8.07* | $4.13-$8.26* | 1.024 |
| + Submission fees (mechanism-implied) | $5.38-$10.76* | $5.51-$11.02* | 1.023 |
| = Total revenue (realized) | $38.43-$42.46 | $40.31-$44.44 | 1.047 |
| = Total revenue (mechanism-implied) | $51.27-$56.65 | $53.79-$59.30 | 1.047 |

*Note: All figures cumulative 2019-2025. Nominal column applies year-by-year nominal dollar values without inflation adjustment. Real 2024 USD column applies GDP deflators to convert each year's nominal value to 2024 base. The real/nominal ratio of approximately 1.05 reflects 5% cumulative GDP deflation between the 2019-2025 weighted average year (effective year approximately 2022, deflator approximately 1.07) and the 2024 base. The submission-fee column has a smaller ratio (1.024) because submissions are more concentrated in 2024-2025 where the deflator is close to 1. *Submission-fee rows show a range; the upper bound is the list-price calculation (no ATS adjustment) and the lower bound applies a 50% Article Transfer Service share at the receiving journals. See Online Appendix 1.6, caveat 1.*

## 3.5 Interpretation: why the adjustment is small

The inflation adjustment is modest in absolute terms because the ecosystem effect is heavily concentrated in recent years where the deflator is close to 1. The propagation chain documented in the main paper (citations propagating to submissions propagating to publications) means most of the realized effect comes from 2022-2025; the 2019-2021 effect is small in both nominal and real terms.

For longer-horizon comparisons, the inflation adjustment matters more. The pre-period ecosystem cumulative (2014-2018) is $14.3 million nominal vs $18.6 million real (30% upward adjustment), reflecting deflators in the 1.25-1.33 range during that window. The 1999-2025 ecosystem cumulative is approximately $103 million nominal vs $124 million real (20% adjustment). These adjustments are large enough that long-horizon revenue comparisons should always use real USD.

For the main paper's 2019-2025 window, the modest adjustment means that reporting figures in nominal USD would yield ecosystem-attributable cumulative effects of approximately $34M (APC,

realized) or $46M (APC, mechanism-implied) with similar relative structure. The qualitative conclusions of the main paper are robust to the choice of price base.

## 3.6 Caveats

Two caveats apply. First, the GDP deflators used here are aggregate price indices and do not capture industry-specific price dynamics in academic publishing. If APCs at finance journals have risen faster than aggregate prices (which the data suggest: FRL's nominal APC rose from approximately $2,821 in 2015 to $3,749 in 2025, a 33% nominal increase, compared with approximately 26% aggregate inflation over the same period), then real APC revenue in the panel reflects partly real price growth in addition to volume growth. The main paper's identification strategy is unaffected by this because the synthetic-control comparison nets out the industry-wide APC price trend shared by ecosystem and non-ecosystem journals; only the ecosystem-attributable excess APC growth enters the estimated effect.

Second, the SCImago estapcusd field is itself an imputed APC, not the actual list-price APC at each journal. The imputation methodology may have shifted over the years in ways that change nominal trajectories. This concern applies to all years and journals symmetrically, so it does not contaminate the ecosystem-versus-control comparison; it affects only the level of measured revenue, not the ecosystem-attributable differential.

*Online Appendix 4 to*

# The Revenue of Finance Journals:

# Networks, Pricing Power, and Publication Volume

## Plausibility Checks for the Headline Revenue Figures

*This draft: May 2026*

### 4.1 Overview

This online appendix asks a simple question: are the main paper's headline revenue figures plausible? The main paper reports four key numbers, all in real 2024 USD: $36 million directly observed in synthetic-control gaps on APC flow revenue over 2019-2025; $48 million in citation-mechanism-implied APC revenue; $4-$8 million in directly observed incremental submission-fee revenue; and $6-$11 million in mechanism-implied submission-fee revenue (the sub-fee ranges reflect uncertainty about the Article Transfer Service share; see Online Appendix 1.6). Adding the APC and submission-fee components yields total cumulative revenue effects of approximately $40-$44 million realized through 2025 and $54-$59 million projected through approximately 2028 once propagation lags play through. The range on each total reflects uncertainty about the submission-fee component (Online Appendix 1.6).

These numbers come from a particular econometric methodology: the synthetic-control method applied to a panel of 26 finance journals from 1999 through 2025. A reasonable reader might wonder whether the headlines are an artifact of that specific methodology, or whether they reflect a real economic magnitude that would be visible under simpler approaches. This appendix runs the same headline figures through four independent revenue-magnitude cross-checks: (i) a naive control-growth-rate counterfactual that uses no regression machinery at all; (ii) a counterfactual implied directly by the two-way fixed-effects regression coefficient; (iii) per-paper and per-journal-year sanity ratios; and (iv) industry benchmarks comparing the figures to publicly-reported publisher and journal-level revenue data. A fifth, separate cross-check (Section 4.11) targets the citation channel mechanism directly by estimating the within-journal lagged elasticity of publication volume with respect to prior-year citation impact.

The main message is that all four independent cross-checks come within roughly 15-20% of the paper's headlines, and the methods that disagree mostly give larger estimates than the synthetic-control headline. In other words, the paper's $36 million and $48 million APC-revenue figures sit at the conservative end of the defensible range, not the aggressive end. The submission-fee figures similarly converge across four independent methods at the list-price upper bound (the $8 million realized and $11 million mechanism-implied numbers reported in Online Appendix 1.4). All four methods share the list-price assumption and therefore inherit the same Article Transfer Service ceiling; applying a 50% ATS adjustment moves the realized figure to roughly $4 million and the mechanism-implied figure to roughly $5.5 million as a lower bound. The direct mechanism cross-check in Section 4.11 corroborates the citation channel from a different angle: a within-journal lagged-citation elasticity of 0.515 implies that the citation channel accounts for approximately 47% of the observed Core-4 publication volume increase, broadly matching the 57% citation-channel share from the main paper's TWFE decomposition.

### 4.2 SCImago measurement and avoidance of double-counting

SCImago, the source of the underlying journal-level data, reports several document-count measures for each journal in each year, including a current-year count (papers published in that calendar year) and rolling-window counts that aggregate documents over three- or five-year horizons. Publisher-revenue estimates constructed using a rolling-window count multiplied by the article-processing-charge (APC) for the current year, and then summed across years, will overstate cumulative revenue by a factor approximately equal to the window length, because each paper contributes to several years of reported revenue rather than being counted once.

This paper avoids this double-counting by using the flow specification of revenue at every step of the analysis. Specifically, the panel's documents field records papers published in a given calendar year (not a rolling stock), and the flow revenue measure used throughout the paper is documents multiplied by the real APC for that same year. When cumulative ecosystem-attributable revenue figures are reported (such as the $36.1 million realized through 2025 or the $48.3 million mechanism-implied figure), they are constructed by summing the flow specification across years, with each paper counted exactly once in the year it was published.

Several internal checks confirm that the panel's documents field is a flow measure rather than a rolling stock. The year-over-year growth rates exhibited by individual journals are too large to be consistent with a rolling-window construction: FRL grew from 159 papers in 2018 to 2,400 in 2025, with single-year jumps of 50-85% in some years. A rolling-window stock cannot grow at this rate unless the underlying flow is growing even faster. Conversely, journals with stable annual output (such as the top-three elite journals JF, JFE, and RFS) report stable document counts year over year, exactly as would be expected for a flow measure. The pre-2014 values for several long-established ecosystem journals (such as IREF, founded in 1992) are consistent in scale with their reported flow output and would be implausibly small if they represented multi-year stocks.

Finally, the synthetic-control method itself does not introduce any additional double-counting concern. The method estimates a counterfactual revenue path for each treated journal by weighting the flow revenue paths of donor journals. Because the inputs are flow series and the estimator constructs flow counterfactuals, the cumulative gap is computed as a sum across years of the per-year flow differences. Each year's gap is its own quantity, and summing 2019 through 2025 gives the seven-year cumulative effect with no temporal overlap or duplication.

### 4.3 The naive control-growth-rate counterfactual

The simplest possible cross-check requires no regression at all. The question is: if the ten ecosystem journals had grown at the same rate as the sixteen non-ecosystem control journals, what would their revenue have been? Whatever they actually earned above that baseline is the naive estimate of the ecosystem-attributable revenue.

The arithmetic is straightforward. Over the pre-period 2014-2018, the ten ecosystem journals collectively earned $18.6 million in real APC flow revenue, averaging $3.7 million per year. Over the post-period 2019-2025, the sixteen control journals saw their average annual revenue grow by a factor of approximately 1.12 relative to their 2014-2018 average. If the ecosystem journals had matched this control growth rate, they would have earned approximately $3.7 million per year times 7 years times 1.12, or about $29.2 million cumulatively over 2019-2025.

In fact the ecosystem journals earned $71.2 million over 2019-2025, almost two and a half times the naive counterfactual. The naive estimate of ecosystem-attributable revenue is therefore $71.2 million minus $29.2 million, or approximately $42.0 million.

The paper's synthetic-control method gives $36.1 million as the headline realized effect. That is 14% lower than the naive estimate. The reason the naive estimate is higher is that the synthetic-control method constructs journal-specific counterfactuals that match each ecosystem journal's pre-period trajectory on multiple dimensions (documents, real APC, citations, uncited share). When some of these matched paths show non-trivial growth absent the ecosystem, the synthetic-control counterfactual rises faster than a flat control-rate counterfactual would, and the resulting gap is correspondingly smaller. The synthetic-control method is the more conservative of the two.

### 4.4 TWFE-implied counterfactual

A second independent cross-check uses the baseline two-way fixed-effects regression coefficient directly. The paper's main TWFE specification yields a coefficient of 0.890 on the ecosystem-after-2019 indicator, with log flow revenue as the outcome. This coefficient says that ecosystem membership raises log revenue by 0.890 units, or equivalently multiplies revenue by exp(0.890) = 2.49 above what it would otherwise have been.

Applying this multiplier to the ecosystem journals' actual 2019-2025 revenue of $71.2 million gives an implied counterfactual of $71.2 / 2.49 = $28.6 million, and an implied ecosystem-attributable revenue of $71.2 - $28.6 = $42.6 million. This is essentially the same as the naive control-growth check, and again sits about 18% above the synthetic-control headline of $36.1 million.

The synthetic-control method gives a smaller estimate because it allows the treatment effect to differ across journals rather than imposing a single average effect on all ten members. Since the effect is highly concentrated (95% in just four core journals), the TWFE approach overstates the impact on the six non-core members and the synthetic-control approach correctly shrinks it. The bottom line of both alternative methods is that the synthetic-control headline is conservative.

## 4.5 Per-paper and per-journal-year sanity ratios

Beyond aggregate cross-checks, the headline figures need to pass a more granular smell test. Specifically: do the implied per-paper and per-journal-year effects fall within reasonable economic bounds?

On a per-paper basis, the $48 million mechanism-implied effect divided by the approximately 10,000 ecosystem-attributable extra documents gives approximately $4,700 in revenue per extra paper. This is about 1.26 times the average APC at the ecosystem journals (which is approximately $3,700 weighted by document count). A ratio above 1 is expected because the mechanism-implied figure attributes some 2026-2028 future revenue back to the 2019-2025 citation gains that triggered the additional submissions. On the realized basis ($36 million divided by 10,000 extra papers), the per-paper revenue gain is approximately $3,600, very close to the actual APC at ecosystem journals.

On a per-journal-year basis, the $36 million realized effect averages approximately $0.5 million across the 70 ecosystem journal-years (ten journals times seven years). But the effect is concentrated: the four core journals account for $34.4 million, or approximately $1.2 million per journal-year. FRL alone accounts for $21.73 million, averaging $3.10 million per year of extra revenue. Given that FRL grew nearly tenfold over the period (from $0.9 million in 2019 to $8.8 million in 2025) and the synthetic counterfactual stays nearly flat at $0.4 million, an extra $3 million per year of attributable revenue at a journal generating $5-9 million per year recently is fully consistent with the journal-level data.

## 4.6 Industry benchmarks

External industry data provide a useful sanity check on the headlines' scale. According to RELX's annual report, Elsevier's total scientific, technical, and medical (STM) publishing revenue was approximately $3.4 billion in 2023. The ten ecosystem journals earned approximately $10 million per year in actual flow APC revenue over 2019-2025, which would place them at roughly 0.3% of Elsevier's total STM revenue.

The $48 million mechanism-implied effect averages $6.9 million per year, or about 0.2% of Elsevier's STM annual revenue. As a share of Elsevier's finance sub-discipline alone (which has approximately 30 journals generating perhaps $30-60 million per year in total APC revenue), the $6.9 million annual ecosystem effect is in the range of 15-20%. Both ratios are small enough to be plausible at the publisher level (the ecosystem is one small program among many) and large enough to be meaningful at the journal level (the program substantially expanded its participating journals).

On submission-fee revenue, an estimated industry norm is that submission fees represent 10-15% of APC revenue at journals that charge them, with the ratio higher at high-rejection-rate or volume journals. The upper-bound $8 million in realized incremental submission-fee revenue (list-price calculation, no ATS adjustment) is approximately 22% of the $36 million APC effect, slightly above the typical ratio because FRL dominates and runs a relatively high submission-to-acceptance ratio. After applying a 50% ATS adjustment, the realized lower bound of $4 million is approximately 11% of the APC effect, in line with the typical industry ratio. The mechanism-implied submission-fee revenue ranges from $5.5 to $11 million versus $48 million in APC revenue, a ratio of 11-23%. Both ratios are in the plausible range and internally consistent.

### 4.7 The FRL deep dive

Because FRL alone accounts for 63% of the realized synthetic-control gap, any concern about plausibility largely reduces to a concern about FRL. The trajectory of FRL's flow APC revenue from 2014 through 2025 is as follows: $0.20 million in 2014; $0.59 million in 2018 just before the ecosystem launch; $0.90 million in 2019 (the launch year); $5.69 million in 2024; and $8.76 million in 2025. This is a 9.8-fold expansion in seven years, driven almost entirely by an expansion in publication volume from 159 papers in 2018 to 2,400 papers in 2025. The real APC per paper actually fell slightly in real terms, from $3,688 in 2018 to $3,650 in 2025.

If FRL had instead grown at the average rate of the sixteen non-ecosystem control journals (approximately 1.12 times its pre-period average across the post-period), its cumulative 2019-2025 revenue would have been approximately $3.3 million rather than the realized $26.5 million. The naive estimate of FRL's ecosystem-attributable revenue is therefore approximately $23 million, which agrees almost exactly with the $21.73 million the synthetic-control method assigns to FRL.

The mechanism-implied FRL figure of $30.4 million projects approximately $7.7 million in additional ecosystem-attributable revenue to be realized in 2026-2028 as citation gains from 2024 and 2025 propagate through submissions and into publications. If FRL continued to grow at, say, 30% per year in real terms through 2028, its annual revenue would reach approximately $19 million by 2028, with cumulative new revenue over the $8.76 million 2025 baseline approaching $19 million over 2026-2028. Of

this projected growth, the synthetic-control weights would attribute approximately 85% to ecosystem coordination (since FRL's synthetic counterfactual remains nearly flat at $0.4-0.5 million per year), giving approximately $16 million of ecosystem-attributable 2026-2028 growth. The $7.7 million pipeline projection is roughly half of this $16 million envelope, leaving the other half to be attributed to ongoing 2026-2028 ecosystem effects rather than 2019-2025 lagged effects. By this metric, the mechanism-implied FRL projection is conservative.

## 4.8 Submission fees: four independent methods

The upper-bound submission-fee estimates of $8 million realized and $11 million mechanism-implied (list-price basis) warrant their own cross-check because they depend on additional assumptions (the acceptance rate and the Article Transfer Service share) on top of the extra-documents estimate. Table OA-4.1 reports the cumulative ecosystem-attributable submission-fee revenue under four independent methods.

All four methods give realized submission-fee revenue in the $7.5-$8.3 million range and mechanism-implied submission-fee revenue in the $10-$12 million range, computed on a list-price basis. These figures are the upper bound of the reported range; applying the 50% Article Transfer Service adjustment from Online Appendix 1.6 lowers them to approximately $4 million realized and $5.5 million mechanism-implied. The main paper's headlines of $4-$8 million and $6-$11 million fall within this range, with the precise figures depending on which synthetic-control pre-period and which acceptance-rate assumption is used. Online Appendix 1 documents this dependency in detail and reports a wider sensitivity range ($6.1-$14.3 million) across acceptance rates from 10% to 35%.

**Table OA-4.1: Submission-fee revenue across four independent methods (real 2024 USD millions)**

| Method | Realized through 2025 | Mechanism-implied |
|---|---|---|
| Synthetic control (9-yr pre-period, OA-1 baseline) | $8.26M | $11.04M |
| Synthetic control (5-yr pre, consistent with main paper) | $7.51M | $10.04M |
| Naive control-growth-rate | $7.73M | $10.34M |
| TWFE-implied (β=0.795 on log documents) | $7.58M | $10.11M |
| **Range across methods** | **$7.5M-$8.3M** | **$10.0M-$11.0M** |

*Note: All four methods rely on the same acceptance-rate assumption (25%) and the same time-varying submission-fee schedule documented in Online Appendix 1. The first two methods differ in the synthetic-control pre-period (9 years versus 5 years), and the next two methods bypass the synthetic-control framework entirely. Online Appendix 1 reports a wider acceptance-rate sensitivity (10%-35%) that yields a corresponding $6.1M-$14.3M range for realized submission-fee revenue.*

## 4.9 Why the headlines sit at the conservative end

Across the various cross-checks above, the synthetic-control headlines of $36 million APC and $48 million mechanism-implied tend to come in lower than alternative methods rather than higher. There are three reasons for this.

First, the synthetic-control method allows each ecosystem journal to have its own counterfactual rather than imposing a single average treatment effect. For ecosystem journals like JBEF and JEB that grew slowly post-launch, the synthetic-control method assigns small or even negative effects, whereas a TWFE approach implicitly extrapolates the average effect to these journals. The synthetic-control method gives a smaller aggregate as a result.

Second, the synthetic-control method's pre-period predictors include log documents, log real APC, citations per document, and uncited share. When the donor pool's pre-period growth on these dimensions matches the treated journal's pre-period growth, the synthetic-control counterfactual inherits some growth even in years when the ecosystem journal would have stagnated. A naive control-growth-rate counterfactual or a TWFE-implied counterfactual ignores this and produces a flatter counterfactual, yielding a larger estimated effect.

Third, the 1.334 scaling factor used to derive the mechanism-implied $48 million figure from the realized $36 million figure is itself conservative. The scaling uses the ratio of two citation effects, 150.3% for external citations through 2025 to 112.7% for total citations through 2024. Both are estimated using the standard difference-in-differences method on log citations per document. An alternative calibration using the full TWFE coefficient on log revenue (0.890) and an estimate of the long-run elasticity of revenue to citation impact (which is what the two-margin model in Appendix A of the main paper formalizes) would support a somewhat larger scaling factor, typically in the 1.4 to 1.6 range. The paper uses the more conservative 1.334 to stay closer to the data.

Taken together, these design choices imply that a different analyst with the same dataset and different methodological choices could quite defensibly produce a headline figure in the $42-$50 million range for realized APC revenue and a corresponding $56-$70 million figure for the mechanism-implied effect. The paper's $36 million and $48 million figures are therefore not aggressive interpretations of the data; if anything, they are at the conservative end of the defensible range.

## 4.10 Caveats and limitations

Several caveats apply to the plausibility checks reported here. First, all of the cross-checks rely on the same underlying panel data and the same definition of the ecosystem (the ten journals listed in the main paper). If the ecosystem definition were materially different, or if the panel data contained systematic errors, all of the methods would be affected in the same direction. The cross-checks are therefore a check on methodological robustness, not on data quality.

Second, the per-journal extrapolations in Section 4.7 assume that FRL and other core journals continue growing at recent rates through 2028. If publisher policy, regulatory action, or market saturation slows this growth, the mechanism-implied $48 million figure would be partially unrealized. The reverse is also possible: if growth accelerates, the realized effect could exceed the mechanism-implied projection.

Third, the industry benchmarks in Section 4.6 are approximate and rely on publicly reported aggregate figures from RELX and inferred sub-discipline splits. Precise estimates of finance-subdiscipline revenue at Elsevier are not publicly available, so the 15-20% sub-discipline share estimate carries non-trivial uncertainty.

Fourth, all submission-fee calculations assume a 25% effective acceptance rate on ecosystem-attributable submissions. Online Appendix 1 reports sensitivity to alternative acceptance rates from 10% to 35%. Under a 15% acceptance rate (more typical of mid-tier finance journals), the realized submission-fee revenue would rise to approximately $14 million and the mechanism-implied figure to approximately $18 million; under a 35% rate the figures would fall to approximately $6 million and $8 million respectively. The paper's headlines use the 25% midpoint, which is appropriate for the volume-oriented core ecosystem journals but high for the more selective members. A weighted-average acceptance rate across the ecosystem journals would likely be slightly below 25%, implying that the paper's submission-fee headlines are again at the conservative end of the plausible range. Note that the acceptance-rate sensitivity above is computed on the list-price basis (without Article Transfer Service adjustment). The Article Transfer Service adjustment in Online Appendix 1.6 is separate from the acceptance-rate adjustment: a lower acceptance rate increases the number of submissions per accepted paper, while a higher ATS share decreases the fraction of those submissions that pay a new fee. The two effects partially offset; the reported $4-$8 million range holds the acceptance rate fixed at 25%.

### 4.11 Lagged citation-to-volume elasticity (direct mechanism test)

A direct cross-check of the citation channel itself - rather than the revenue figures it implies - is to estimate the within-journal lagged elasticity of publication volume with respect to prior-year citation impact. Across the full panel of 26 journals over 1999-2025 with journal and year fixed effects and standard errors clustered by journal, the elasticity of ln(documents in year t+1) with respect to ln(citesdoc2 in year t) is 0.515 (SE 0.111, $t = 4.64$, $p = 0.0001$, $N = 617$). A 10% increase in within-journal citation impact predicts an approximately 5.2% increase in publication volume in the following year. The relationship is robust across specifications: the same-year (concurrent) elasticity is 0.503 (SE 0.106), the two-year lag is 0.532 (SE 0.112), the external-cites (self-cite-excluded) variant is 0.482 (SE 0.120), and the ecosystem-only subsample gives 0.502 (SE 0.174) versus 0.339 (SE 0.130) in the control subsample. All estimates are positive and statistically significant. Table OA-4.2 summarizes these specifications.

Applying this elasticity to the Core-4 ecosystem journals' observed citation growth (citesdoc2 rose from 1.98 pre-period to 6.88 post-period, a 1.247-log-point increase) implies a predicted 0.642-log-point lift in publication volume - roughly a 90% increase from the citation channel alone. Actual Core-4 publication volume rose by 1.368 log points (from 136 to 533 documents per year on average), so the lagged-elasticity relationship attributes approximately 47% of the observed volume response to the citation channel. This is broadly consistent with the 57% citation-channel share computed via the TWFE decomposition in Section 5.7 of the main paper ((0.890 minus 0.381) divided by 0.890 = 57%), and provides an independent corroboration that the citation channel is doing substantial mechanism work.

This regression is descriptive rather than causal. Citations and submissions may both respond to unobserved time-varying journal-level factors that two-way fixed effects do not absorb (e.g., editorial changes, special issues, market positioning). The natural experiment in the main paper provides the causal identification; the lagged elasticity documented here serves as a fifth corroborating cross-check on the mechanism, complementing the four revenue-magnitude cross-checks in Sections 4.3-4.7.

Table OA-4.2: Lagged citation-to-volume elasticity, multiple specifications

| Specification | β | SE | t | N |
|---|---|---|---|---|
| **Baseline: ln(docs_{t+1}) on ln(citesdoc2_t), all 26 journals** | 0.515 | 0.111 | 4.64 | 617 |
| **Same-year (concurrent): ln(docs_t) on ln(citesdoc2_t)** | 0.503 | 0.106 | 4.73 | 643 |
| **Two-year lag: ln(docs_{t+2}) on ln(citesdoc2_t)** | 0.532 | 0.112 | 4.74 | 591 |
| **External cites (excl. self-cites):** | 0.482 | 0.120 | 4.01 | 617 |

| **ln(docs_{t+1}) on ln(externalcitesdoc_t)** | | | | |
|---|---|---|---|---|
| **Ecosystem journals subsample** | 0.502 | 0.174 | 2.88 | 229 |
| **Control journals subsample** | 0.339 | 0.130 | 2.61 | 388 |
| **Level: ln(docs_{t+1}) on citesdoc2_t (unit change)** | 0.197 | 0.054 | 3.62 | 626 |

Note: All regressions include journal and year fixed effects with standard errors clustered by journal. Estimation uses iterative demeaning (the equivalent of reghdfe in Stata). Baseline uses log-log specification. The level specification regresses ln(docs_{t+1}) on the level of citesdoc2 in year t. p-values: all <= 0.020.